\documentclass[journal]{IEEEtran}
\usepackage[english]{babel}
\usepackage{graphicx}
\usepackage{amsmath}
\usepackage{mathtools}
\usepackage{cite}

\begin{document}

\title{Information Leakage of Correlated Source Coded Sequences over a Channel with an Eavesdropper}
\author{\authorblockN{Reevana Balmahoon and Ling Cheng }\\
\authorblockA{School of Electrical and Information Engineering\\University of the Witwatersrand\\Private Bag 3, Wits. 2050, Johannesburg, South Africa\\
Email: reevana.balmahoon@students.wits.ac.za, ling.cheng@wits.ac.za}}
\maketitle

\begin{abstract}
A new generalised approach for multiple correlated sources over a wiretap network is investigated. A basic model consisting of two correlated sources where each produce a component of the common information is initially investigated. There are several cases that consider wiretapped syndromes on the transmission links and based on these cases a new quantity, the information leakage at the source/s is determined. An interesting feature of the models described in this paper is the information leakage quantification. Shannon's cipher system with eavesdroppers is incorporated into the two correlated sources model to minimize key lengths. These aspects of quantifying information leakage and reducing key lengths using Shannon's cipher system are also considered for a multiple correlated source network approach. A new scheme that incorporates masking using common information combinations to reduce the key lengths is presented and applied to the generalised model for multiple sources. 
\end{abstract}

\section{Introduction}
\begin{figure}[ht]
\centering
\includegraphics [scale = 0.7]{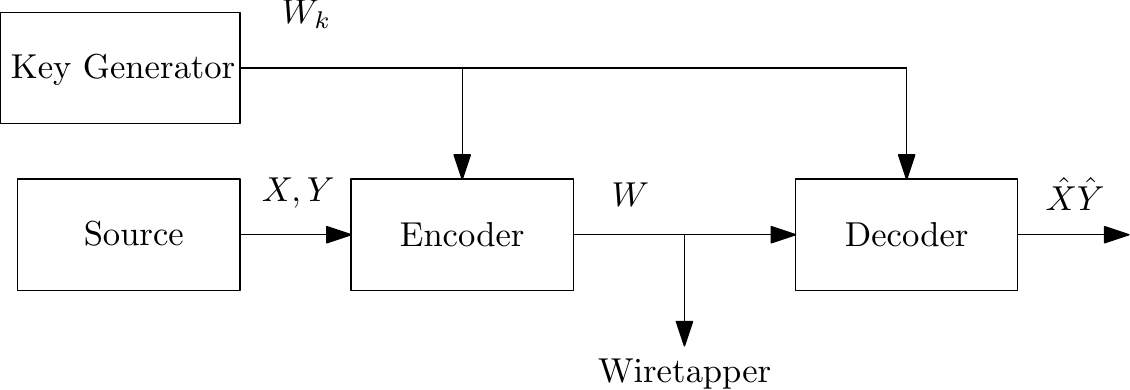}
\caption{Yamamoto's development of the Shannon Cipher System}
\label{fig:yamamoto_shannoncipher}
\end{figure}

Keeping information secure has become a major concern with the advancement in technology. In this work, the information theory aspect of security is analyzed, as entropies are used to measure security. The system also incorporates some traditional ideas surrounding cryptography, namely Shannon's cipher system and adversarial attackers in the form of eavesdroppers. In cryptographic systems, there is usually a message in plaintext that needs to be sent to a receiver. In order to secure it, the plaintext is encrypted so as to prevent eavesdroppers from reading its contents. This ciphertext is then transmitted to the receiver. Shannon's cipher system (mentioned by Yamamoto\cite{shannon1_yamamoto}) incorporates this idea. The definition of Shannon's cipher system has been discussed by Hanawal and Sundaresan~\cite{hanawal_shannon}. In Yamamoto's~\cite{shannon1_yamamoto} development on this model, a correlated source approach is introduced. This gives an interesting view of the problem, and is depicted in Figure~\ref{fig:yamamoto_shannoncipher}. Correlated source coding incorporates the lossless compression of two or more correlated data streams. Correlated sources have the ability to decrease the bandwidth required to transmit and receive messages because a syndrome (compressed form of the original message) is sent across the communication links instead of the original message. A compressed message has more information per bit, and therefore has a higher entropy because the transmitted information is more unpredictable. The unpredictability of the compressed message is also beneficial in terms of securing the information. 

The source sends information for the correlated sources, $X$ and $Y$ along the main transmission channel. A key $W_k$, is produced and used by the encoder when producing the ciphertext. The wiretapper has access to the transmitted codeword, $W$. The decoded codewords are represented by $\widehat{X}$ and $\widehat{Y}$. In Yamamoto's scheme the security level was also focused on and found to be $\frac{1}{K} H(X^K,Y^K|W)$ (i.e. the joint entropy of $X$ and $Y$ given $W$, where $K$ is the length of $X$ and $Y$) when $X$ and $Y$ have equal importance, which is in accordance with traditional Shannon systems where the security is measured by the equivocation. When one source is more important than the other then the security level is measured by the pair of the individual uncertainties $(\frac{1}{K} H(X^K|W), \frac{1}{K} H(Y^K|W))$. 

In practical communication systems links are prone to eavesdropping and as such this work incorporates wiretapped channels, i.e. channels where an eavesdropper is present.

There are specific kinds of wiretapped channels that have been developed. The mathematical model for this Wiretap Channel is given by Rouayheb \textit{et al.}~\cite{ref12_rouayheb_soljanin}, and can be explained as follows: the channel between a transmitter and receiver  is error-free and can transmit $n$ symbols $Y=(y_1,\ldots,y_n)$ from which $\mu$ bits can be observed by the eavesdropper and the maximum secure rate can be shown to equal $n-\mu$ bits. The security aspect of wiretap networks have been looked at in various ways by Cheng \textit{et al.} \cite{ref21_cheng_yeung}, and Cai and Yeung \cite{ref11_cai_yeung}, emphasising that it is of concern to secure these type of channels. 

Villard and Piantanida \cite{pablo_secure_multiterminal} also look at correlated sources and wireap networks: A source sends information to the receiver and an eavesdropper has access to information correlated to the source, which is used as side information. There is a second encoder that sends a compressed version of its own correlation observation of the source privately to the receiver. Here, the authors show that the use of correlation decreases the required communication rate and increases secrecy. Villard \textit{et al.} \cite{pablo_secure_transmission_receivers} explore this side information concept further where security using side information at the receiver and eavesdropper is investigated. Side information is generally used to assist the decoder to determine the transmitted message. An earlier work involving side information is that by Yang \textit{et al.}~\cite{feedback_yang}. The concept can be considered to be generalised in that the side information could represent a source. It is an interesting problem when one source is more important and Hayashi and Yamamoto\cite{Hayashi_coding} consider it in another scheme, where only $X$ is secure against wiretappers and $Y$ must be transmitted to a legitimate receiver. They develop a security criterion based on the number of correct guesses of a wiretapper to retrieve a message. In an extension of the Shannon cipher system, Yamamoto \cite{coding_yamamoto} investigated the secret sharing communication system. 

In this case, we generalise a model for correlated sources across a channel with an eavesdropper and the security aspect is explored by quantifying the information leakage and reducing the key lengths when incorporating Shannon's cipher system. 

This paper initially describes a two correlated source model across wiretapped links, which is detailed in Section II. In Section III, the information leakage is investigated and proven for this two correlated source model. The information leakage is quantified to be the equivocation subtracted from the total obtained uncertainty. In Section IV the two correlated sources model is looked at according to Shannon's cipher system. The notation contained in the tables will be clarified in the following sections. The proofs for this Shannon cipher system aspect are detailed in Section V. Section VI details the extension of the two correlated source model where multiple correlated sources in a network scenario is investigated. There are two subsections here; one quantifying information leakage for the Slepian-Wolf scenario and the other incorporating Shannon's cipher system where key lengths are minimized and a masking method to save on keys is presented. Section VII explains how the models detailed in this paper are a generalised model of Yamamoto's~\cite{shannon1_yamamoto} model, and further offers comparison to other models. The future work for this research is detailed in Section VIII and the paper is concluded in Section IX.

\section{Model}

\begin{figure}[ht]
\centering
\includegraphics [scale = 0.7]{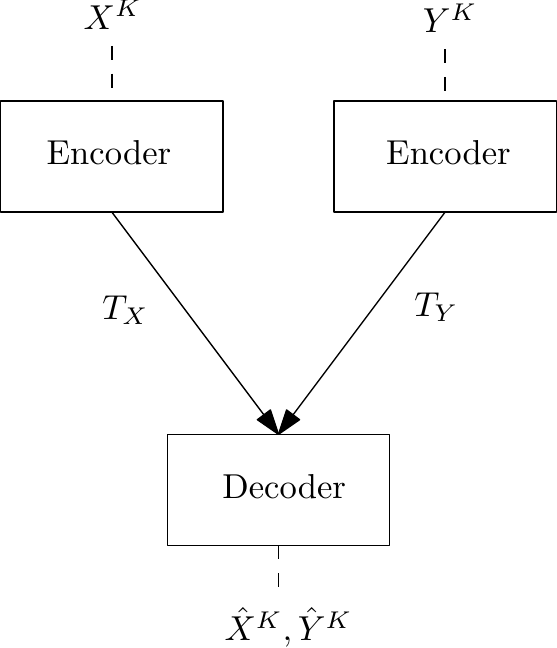}
\caption{Correlated source coding for two sources}
\label{fig:new_model}
\end{figure}

\begin{figure}[ht]
\centering
\includegraphics [scale = 0.7]{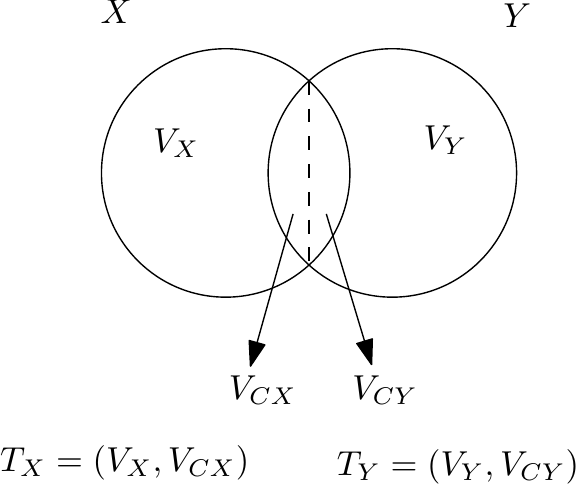}
\caption{The relation between private and common information}
\label{fig:new_venn2}
\end{figure}

The independent, identically distributed (i.i.d.) sources $X$ and $Y$ are mutually correlated random variables, depicted in Figure~\ref{fig:new_model}. The alphabet sets for sources $X$ and $Y$ are represented by $\mathcal{X}$ and $\mathcal{Y}$ respectively. Assume that ($X^K$, $Y^K$) are encoded into two syndromes ($T_{X}$ and $T_{Y}$). We can write $T_X = (V_X, V_{CX})$ and $T_Y = (V_Y, V_{CY})$ where $T_X$ and $T_Y$ are the syndromes of $X$ and $Y$. Here, $T_X$ and $T_Y$ are characterised by $(V_X, V_{CX})$ and $(V_Y, V_{CY})$ respectively. The Venn diagram in Figure \ref{fig:new_venn2}  easily illustrates this idea where it is shown that $V_X$ and $V_Y$ represent the private information of sources $X$ and $Y$ respectively and $V_{CX}$ and $V_{CY}$ represent the common information between $X^K$ and $Y^K$ generated by $X^K$ and $Y^K$ respectively.

The correlated sources $X$ and $Y$ transmit messages (in the form of syndromes) to the receiver along wiretapped links. The decoder determines $X$ and $Y$ only after receiving all of $T_X$ and $T_Y$. The common information between the sources are transmitted through the portions $V_{CX}$ and $V_{CY}$. In order to decode a transmitted message, a source's private information and both common information portions are necessary. This aids in security as it is not possible to determine, for example $X$ by wiretapping all the contents transmitted along $X$'s channel only. This is different to Yamamoto's~\cite{shannon1_yamamoto} model as here the common information consists of two portions.  The aim is to keep the system as secure as possible and these following sections show how it is achieved by this new model. 

We assume that the function $F$ is a one-to-one process with high probability, which means based on $T_X$ and $T_Y$ we can retrieve $X^K$ and $Y^K$ with minimal error. Furthermore, it reaches the Slepian-Wolf bound, $H(T_X, T_Y)=H(X^K,Y^K)$. Here, we note that the lengths of $T_X$ and $T_Y$ are not fixed, as it depends on the encoding process and nature of the Slepian-Wolf codes. The process is therefore not ideally one-to-one and reversible and is another difference between our model and Yamamoto's~\cite{shannon1_yamamoto} model.

The code described in this section satisfies the following inequalities for $\delta > 0$ and sufficiently large $K$.

\begin{eqnarray}
Pr \{X^K \neq G(V_X, V_{CX}, V_{CY})\} \le \delta
\label{x_prob}
\end{eqnarray}

\begin{eqnarray}
Pr \{Y^K \neq G(V_Y, V_{CX}, V_{CY})\} \le \delta
\label{y_prob}
\end{eqnarray}

\begin{eqnarray}
H(V_X, V_{CX}, V_{CY})\le H(X^K) + \delta 
\label{x_entropy}
\end{eqnarray}

\begin{eqnarray}
H(V_Y, V_{CX}, V_{CY})\le H(Y^K) + \delta 
\label{y_entropy}
\end{eqnarray}

\begin{eqnarray}
H(V_X, V_Y, V_{CX}, V_{CY})\le H(X^K,Y^K) + \delta 
\label{xy_entropy}
\end{eqnarray}

\begin{eqnarray}
H(X^K|V_X, V_Y) \geq H(V_{CX}) + H(V_{CY}) - \delta 
\label{H_inequality1}
\end{eqnarray}

\begin{eqnarray}
H(X^K|V_{CX}, V_{CY}) \geq H(V_X) + H(V_{CY}) - \delta 
\label{H_inequality2}
\end{eqnarray}

\begin{eqnarray}
H(X^K|V_{CX}, V_{CY}, V_Y) \geq H(V_X) + H(V_{CY}) - \delta 
\label{H_inequality3}
\end{eqnarray}

\begin{eqnarray}
H(V_{CX}) + H(V_X) - \delta \le H(X^K|V_{CY}, V_{Y})  \nonumber
\\ \le H(X) - H(V_{CY}) + \delta
\label{H_inequality4}
\end{eqnarray}
\\
where $G$ is a function to define the decoding process at the receiver. It can intuitively be seen from \eqref{x_entropy} and \eqref{y_entropy} that $X$ and $Y$ are recovered from the corresponding private information and the common information produced by $X^K$ and $Y^K$. Equations \eqref{x_entropy}, \eqref{y_entropy} and \eqref{xy_entropy} show that the private information and common information produced by each source should contain no redundancy. 
It is also seen from \eqref{H_inequality2} and \eqref{H_inequality3} that $V_Y$ is independent of $X^K$ asymptotically. Here, $V_X$, $V_Y$, $V_{CX}$ and $V_{CY}$ are disjoint, which ensures that there is no redundant information sent to the decoder. 

To recover $X$ the following components are necessary: $V_X$, $V_{CX}$ and $V_{CY}$. This comes from the property that $X^K$ cannot be derived from $V_X$ and $V_{CX}$ only and part of the common information between $X^K$ and $Y^K$ is produced by $Y^K$.

Yamamoto~\cite{shannon1_yamamoto} proved that a common information between $X^K$ and $Y^K$ is represented by the mutual information $I(X;Y)$. Yamamoto~\cite{shannon1_yamamoto} also defined two kinds of common information. The first common information is defined as the rate of the attainable minimum core $V_C$ (i.e. $V_{CX}, V_{CY}$ in this model) by removing each private information, which is independent of the other information, from ($X^K$, $Y^K$) as much as possible. The second common information is defined as the rate of the attainable maximum core $V_C$ such that if we lose $V_C$ then the uncertainty of $X$ and $Y$ becomes $H(V_C)$. Here, we consider the common information that $V_{CX}$ and $V_{CY}$ represent.

We begin demonstrating the relationship between the common information portions by constructing the prototype code ($W_X$, $W_Y$, $W_{CX}$, $W_{CY}$) as per Lemma 1. 

\textit{Lemma 1: For any $\epsilon_0 \geq 0$ and sufficiently large $K$, there exits a code $W_X = F_X(X^K)$, $W_Y = F_Y(Y^K)$, $W_{CX} = F_{CX}(X^K)$, $W_{CY} = F_{CY}(Y^K)$, $\widehat{X}^K,\widehat{Y}^K = G(W_X, W_Y, W_{CX}, W_{CY})$, where $W_X \in I_{M_X}$, $W_Y \in I_{M_Y}$, $W_{CX} \in I_{M_{CX}}$, $W_{CY} \in I_{M_{CY}}$ for $I_{M_{\alpha}}$, which is defined as $\{0, 1, \ldots, M_{\alpha} - 1\}$, that satisfies},

\begin{eqnarray}
Pr\{\widehat{X}^K, \widehat{Y}^K \neq X^K, Y^K\} \le \epsilon
\label{lemma1_1}
\end{eqnarray}

\begin{eqnarray}
H(X|Y) - \epsilon_0 \le \frac{1}{K} H(W_X) \le \frac{1}{K} \log M_X \le H(X|Y) + \epsilon_0
\label{lemma1_2}
\end{eqnarray}

\begin{eqnarray}
H(Y|X) - \epsilon_0 \le \frac{1}{K} H(W_Y) \le \frac{1}{K} \log M_Y \le H(Y|X) + \epsilon_0
\label{lemma1_3}
\end{eqnarray}

\begin{eqnarray}
& & I(X;Y) - \epsilon_0 \le \frac{1}{K} (H(W_{CX}) + H(W_{CY})) \nonumber \\
    & \le & \frac{1}{K} (\log M_{CX} + \log M_{CY}) \le I(X;Y) + \epsilon_0
\label{lemma1_4}
\end{eqnarray}

\begin{eqnarray}
\frac{1}{K} H(X^K|W_Y) \geq H(X) - \epsilon_0
\label{lemma1_5}
\end{eqnarray}

\begin{eqnarray}
\frac{1}{K} H(Y^K|W_X) \geq H(Y) - \epsilon_0
\label{lemma1_6}
\end{eqnarray}

We can see that \eqref{lemma1_2} - \eqref{lemma1_4} mean
\begin{eqnarray}
&& H(X,Y) - 3\epsilon_0 \le \frac{1}{K} (H(W_X) + H(W_Y) + H(W_{CX}) \nonumber \\
& + & H(W_{CY})) \nonumber \\
& \le & H(X,Y) + 3\epsilon_0
\label{lemma1_7}
\end{eqnarray}

Hence from \eqref{lemma1_1}, \eqref{lemma1_7} and the ordinary source coding theorem, ($W_X$, $W_Y$, $W_{CX}$, $W_{CY}$) have no redundancy for sufficiently small $\epsilon_0 \geq 0$. It can also be seen that $W_X$ and $W_Y$ are independent of $Y^K$ and $X^K$ respectively. 

\begin{proof}[Proof of Lemma 1]

As seen by Slepian and Wolf, mentioned by Yamamoto\cite{shannon1_yamamoto} there exist $M_X$ codes for the $P_{Y|X}(y|x)$ DMC (discrete memoryless channel) and $M_Y$ codes for the $P_{X|Y}(x|y)$ DMC. The codeword sets exist as $C^X_i$ and $C^Y_j$, where $C^X_i$ is a subset of the typical sequence of $X^K$ and $C^Y_j$ is a subset of the typical sequence of $Y^K$.
The encoding functions are similar, but we have created one decoding function as there is one decoder at the receiver:

\begin{eqnarray}
f_{Xi}:I_{M_{CX}} \rightarrow C^X_i
\label{lemma1proof_1}
\end{eqnarray}

\begin{eqnarray}
f_{Yj}:I_{M_{CY}} \rightarrow C^Y_j
\label{lemma1proof_1}
\end{eqnarray}

\begin{eqnarray}
g: X^K, Y^K \rightarrow I_{M_{CX}} \times I_{M_{CY}}
\label{lemma1proof_2}
\end{eqnarray}

The relations for $M_X$, $M_Y$ and the common information remain the same as per Yamamoto's and will therefore not be proven here. 

In this scheme, we use the average $(V_{CX}, V_{X},V_{CY}, V_{Y})$ transmitted for many codewords from $X$ and $Y$. Thus, at any time either $V_{CX}$ or $V_{CY}$ is transmitted. Over time, the split between which common information portion is transmitted is determined and the protocol is prearranged accordingly. Therefore all the common information is either transmitted as $l$ or $m$, and as such Yamamoto's encoding and decoding method may be used. 

As per Yamamoto's method the code does exist and that $W_X$ and $W_Y$ are independent of $Y$ and $X$ respectively, as shown by Yamamoto\cite{shannon1_yamamoto}.

\end{proof}

The common information is important in this model as the sum of $V_{CX}$ and $V_{CY}$ represent a common information between the sources. The following theorem holds for this common information:
\\
\textit{Theorem 1:}
\begin{eqnarray}
\frac{1}{K} [H (V_{CX}) + H (V_{CY})] = I (X;Y) 
\label{theorem1}
\end{eqnarray}

where $V_{CX}$ is the common portion between $X$ and $Y$ produced by $X^K$ and $V_{CY}$ is the common portion between $X$ and $Y$ produced by $Y^K$. It is noted that the \eqref{theorem1} holds asymptotically, and does not hold with equality when $K$ is finite. Here, we show the approximation when $K$ is infinitely large.
The private portions for $X^K$ and $Y^K$ are represented as $V_X$ and $V_Y$ respectively. As explained in Yamamoto's~\cite{shannon1_yamamoto} Theorem 1, two types of common information exist (the first is represented by $I(X;Y)$ and the second by $\text{min} (H(X^K), H(Y^K))$. We will develop part of this idea to show that the sum of the common information portions produced by $X^K$ and $Y^K$ in this new model is represented by the mutual information between the sources. 

\begin{proof}[Proof of Theorem 1]
The first part is to prove that $H(V_{CX}) + H(V_{CY}) \geq I(X;Y)$, and is done as follows. 
We weaken the conditions \eqref{x_prob} and \eqref{y_prob} to
\begin{eqnarray}
\text{Pr }\{X^K,Y^K \neq G_{XY} (V_X, V_Y, V_{CX}, V_{CY}\}) \le \delta_1
\label{weakenedxy_prob}
\end{eqnarray}

For any ($V_X$,$V_Y$, $V_{CX}$, $V_{CY}$) $\in C(3\epsilon_0)$ (which can be seen from \eqref{lemma1_7}), we have from \eqref{weakenedxy_prob} and the ordinary source coding theorem that

\begin{eqnarray}
H(X^K,Y^K) - \delta_1 &\le & \frac{1}{K} H(V_X, V_Y, V_{CX}, V_{CY}) \nonumber \\
	& \le & \frac{1}{K} [H(V_X) + H(V_Y) + H (V_{CX})  \nonumber \\
	& + & H(V_{CY})]
\label{theorem1proof_1}
\end{eqnarray}

where $\delta_1 \rightarrow 0$ as $\delta \rightarrow 0$. From Lemma 1,
\begin{eqnarray}
\frac{1}{K} H(V_Y|X^K) \geq \frac{1}{K} H(V_Y) - \delta
\label{theorem1proof_2}
\end{eqnarray}

\begin{eqnarray}
\frac{1}{K} H(V_X|Y^K) \geq \frac{1}{K} H(V_X) - \delta
\label{theorem1proof_3}
\end{eqnarray}

From \eqref{theorem1proof_1} - \eqref{theorem1proof_3},
\begin{eqnarray}
\frac{1}{K} [H(V_{CX}) + H(V_{CY})] &\ge & H(X,Y) - \frac{1}{K} H(V_X)  \nonumber \\
& - & \frac{1}{K} H(V_Y)  - \delta_1							\nonumber \\
& \geq & H(X,Y) - \frac{1}{K} H(V_X|Y)  		\nonumber \\
& - & \frac{1}{K} H(V_Y|X) - \delta_1 - 2\delta
\label{theorem1proof_4}
\end{eqnarray}

On the other hand, we can see that
\begin{eqnarray}
\frac{1}{K} H(X^K, V_Y) \le  H(X,Y) + \delta
\label{theorem1proof_5}
\end{eqnarray}

This implies that 
\begin{eqnarray}
\frac{1}{K} H(V_Y|X^K) \le H(Y|X) + \delta  
\label{theorem1proof_6}
\end{eqnarray}
and 
\begin{eqnarray}
\frac{1}{K} H(V_X|Y^K) \le H(X|Y) + \delta 
\label{theorem1proof_7}
\end{eqnarray}

From \eqref{theorem1proof_4}, \eqref{theorem1proof_6} and \eqref{theorem1proof_7} we get
\begin{eqnarray}
\frac{1}{K} [H(V_{CX}) + H(V_{CY})] & \ge & H(X,Y) - H(X|Y) - H(Y|X)  \nonumber \\
& - & \delta_1 - 4\delta	\nonumber \\
& = & I(X;Y) - \delta_1 - 4\delta
\label{theorem1proof_8}
\end{eqnarray}

It is possible to see from \eqref{lemma1_4} that $H(V_{CX}) +  H(V_{CY}) \le I(X;Y)$. From this result,  \eqref{lemma1proof_2} and \eqref{theorem1proof_8}, and as $\delta_1 \rightarrow 0$ and $\delta \rightarrow 0$ it can be seen that
\begin{eqnarray}
\frac{1}{K} [H (V_{CX} + H(V_{CY})] = I(X;Y)
\end{eqnarray}
\end{proof}
 
This model can cater for a scenario where a particular source, say $X$ needs to be more secure than $Y$ (possibly because of eavesdropping on the $X$ channel). In such a case, the $\frac{1}{K} H(V_{CX})$ term in \eqref{theorem1proof_8} needs to be as high as possible. When this uncertainty is increased then the security of $X$ is increased. Another security measure that this model incorporates is that $X$ cannot be determined from wiretapping only $X$'s link.

\section{Information Leakage}
In order to determine the security of the system, a measure for the amount of information leaked has been developed. This is a new notation and quantification, which emphasizes the novelty of this work. The obtained information and total uncertainty are used to determine the leaked information. Information leakage is indicated by $L_{\mathcal{Q}}^\mathcal{P}$. Here $\mathcal{P}$ indicates the source/s for which information leakage is being quantified, $\mathcal{P} = \{S_1, \ldots, S_n\}$ where $n$ is the number of sources (in this case, $n = 2$). Further, $\mathcal{Q}$ indicates the syndrome portion that has been wiretapped, $\mathcal{Q} = \{V_1, \ldots, V_m\}$ where $m$ is the number of codewords (in this case, $m = 4$).

The information leakage bounds are as follows:
\begin{eqnarray}
L_{V_X,V_Y}^{X^K} \le H(X^K) - H(V_{CX}) - H(V_{CY}) + \delta
\label{L_inequality1}
\end{eqnarray}

\begin{eqnarray}
L_{V_{CX},V_{CY}}^{X^K} \le H(X^K) - H(V_X) - H(V_{CY}) + \delta
\label{L_inequality2}
\end{eqnarray}

\begin{eqnarray}
L_{V_{CX},V_{CY},V_Y}^{X^K} \le H(X^K) - H(V_X) - H(V_{CY}) + \delta
\label{L_inequality3}
\end{eqnarray}

\begin{eqnarray}
&& H(V_{CY}) - \delta \le L_{V_{Y},V_{CY}}^{X^K} \nonumber
\\ & \le & H(X^K) - H(V_{CX}) - H(V_X) + \delta 
\label{L_inequality4}
\end{eqnarray}
 
Here, $V_Y$ is private information of source $Y^K$ and is independent of $X^K$
and therefore does not leak any information about $X^K$, shown
in \eqref{L_inequality2} and \eqref{L_inequality3}. Equation \eqref{L_inequality4} gives an indication of the minimum and maximum amount of leaked information for the interesting case where a syndrome has been wiretapped and its information leakage on the alternate source is quantified. The outstanding common information component is the maximum information that can be leaked. For this case, the common information $V_{CX}$ and $V_{CY}$ can thus consist of added
protection to reduce the amount of information leaked. These bounds developed in \eqref{L_inequality1} - \eqref{L_inequality4} are proven in the next section.

The proofs for the above mentioned information leakage inequalities are now detailed. First, the inequalities in \eqref{H_inequality1} - \eqref{H_inequality4} will be proven, so as to prove that the information leakage equations hold. \\
\begin{proof}[Lemma 2]
The code ($V_X$, $V_{CX}$, $V_{CY}$, $V_Y$) defined at the beginning of Section I, describing the model and \eqref{x_prob} - \eqref{xy_entropy} satisfy \eqref{H_inequality1} - \eqref{H_inequality4}. Then the information leakage bounds are given by \eqref{L_inequality1} - \eqref{L_inequality4}.
\\\\ \textit{Proof for \eqref{H_inequality1}}:
\begin{eqnarray}
& & \frac{1}{K} H(X^K|V_X,V_Y) 												\nonumber\\
& = & \frac{1}{K} [H(X^K,V_X,V_Y) - H(V_X,V_Y)] 							\nonumber\\
& = & \frac{1}{K} [H(X^K,V_Y) - H(V_X,V_Y)] 								\label{lemma2_ref1}\\
& = & \frac{1}{K} [H(X^K|V_Y) + I(X^K;V_Y) + H(V_Y|X^K)] 							\nonumber\\
& & - \frac{1}{K} [H(V_X|V_Y) + I(V_X;V_Y) + H(V_Y|V_X)] 						\nonumber\\
& = & \frac{1}{K} [H(X^K|V_Y) + H(V_Y|X^K) - H(V_X|V_Y)							\nonumber\\
& & - H(V_Y|V_X)]															\nonumber\\
& = & \frac{1}{K} [H(X^K) + H(V_Y) - H(V_X) - H(V_Y)]							\label{lemma2_ref2}\\ 
& = & \frac{1}{K} [H(X^K) - H(V_X)]											\nonumber\\
& \geq & \frac{1}{K} [H(V_X) + H(V_{CX}) + H(V_{CY}) - H(V_X)]	- \delta			\nonumber\\
& = & \frac{1}{K} [H(V_{CX}) + H(V_{CY})] -\delta					 			
\label{lemma2_part1}
\end{eqnarray}

where \eqref{lemma2_ref1} holds because $V_X$ is a function of $X$ and \eqref{lemma2_ref2} holds because $X$ is independent of $V_Y$ asymptotically and $V_X$ is independent of $V_Y$ asymptotically.

For the proofs of \eqref{H_inequality2} and \eqref{H_inequality3}, the following simplification for $H(X|V_{CY})$ is used:
\begin{eqnarray}
H(X^K|V_{CY}) & = & H(X^K,Y^K) - H(V_{CY}) \nonumber \\
& = & H(X^K) + H(V_{CY}) - I(X; V_{CY}) - H(V_{CY}) \nonumber \\
& = & H(X^K) + H(V_{CY}) - H(V_{CY}) - H(V_{CY}) \nonumber \\
& + & {\delta}_1 \label{new_55} \\
& = & H(X^K) - H(V_{CY}) = {\delta}_1
\label{simpli}
\end{eqnarray}

where $I(X; V_{CY})$ approximately equal to $H(V_{CY})$ in \eqref{new_55} can be seen intuitively from the Venn diagram in Figure \ref{fig:new_venn2}. Since it is an approximation, ${\delta}_1$, which is smaller than $\delta$ in the proofs below has been added to cater for the tolerance. 
\\\\  \textit{Proof for \eqref{H_inequality2}}:
\begin{eqnarray}
& & \frac{1}{K} H(X^K|V_{CX},V_{CY})					\nonumber\\
& = & \frac{1}{K} [H(X^K,V_{CX},V_{CY}) - H(V_{CX},V_{CY})]						\nonumber\\
& = & \frac{1}{K} [H(X^K,V_{CY}) - H(V_{CX},V_{CY})] 								\label{lemma2_ref3}\\
& = & \frac{1}{K} [H(X^K) - H(V_{CY}) + I(X;V_{CY}) + H(V_{CY}|X^K)] 				\nonumber\\
& & - \frac{1}{K} [H(V_{CX}|V_{CY}) + I(V_{CX};V_{CY}) + H(V_{CY}|V_{CX})] \nonumber \\
& + & \delta_1		\nonumber\\	
& = & \frac{1}{K} [H(X^K) - H(V_{CY}) + H(V_{CY})- H(V_{CX}) - H(V_{CY})]	 \nonumber \\
& + & \delta_1					\label{lemma2_ref4} \\
& = & \frac{1}{K} [H(X^K) - H(V_{CY}) - H(V_{CX})]								+ \delta_1			\nonumber\\
& \geq & \frac{1}{K} [H(V_X) + H(V_{CX}) + H(V_{CY}) - H(V_{CY}) - H(V_{CX})] -\delta			\nonumber\\
& = & \frac{1}{K} H(V_X) + \delta_1 -\delta
\label{lemma2_part2}
\end{eqnarray}

where \eqref{lemma2_ref3} holds because $V_{CX}$ is a function of $X^K$ and \eqref{lemma2_ref4} holds because $X$ is independent of $V_{CY}$ asymptotically and $V_{CX}$ is independent of $V_{CY}$ asymptotically. 

The proof for $H(X|V_{CX},V_{CY},V_Y)$ is similar to that for $H(X|V_{CX},V_{CY})$, because $V_Y$ is independent of $X$.
\\\\ \textit{Proof for \eqref{H_inequality3}}:
\begin{eqnarray}
& & \frac{1}{K} H(X^K|V_{CX},V_{CY},V_Y)										\nonumber\\
& = & \frac{1}{K} H(X^K|V_{CX},V_{CY})
\label{lemma2_ref5}\\
& = & \frac{1}{K} [H(X^K,V_{CX},V_{CY}) - H(V_{CX},V_{CY})]						\nonumber\\
& = & \frac{1}{K} [H(X^K,V_{CY}) - H(V_{CX},V_{CY})] 								\label{lemma2_ref6}\\
& = & \frac{1}{K} [H(X^K) - H(V_{CY}) + I(X;V_{CY}) + H(V_{CY}|X^K)] 					\nonumber\\
& & - \frac{1}{K} [H(V_{CX}|V_{CY}) + I(V_{CX};V_{CY}) + H(V_{CY}|V_{CX})]  \nonumber \\
& + & \delta_1		\nonumber\\	
& = & \frac{1}{K} [H(X^K) - H(V_{CY}) + H(V_{CY})- H(V_{CX}) 		\nonumber\\
& - &  H(V_{CY})] + \delta_1				\label{lemma2_ref7}\\
& = & \frac{1}{K} [H(X^K) - H(V_{CY}) - H(V_{CX})]								+ \delta_1			\nonumber\\
& \geq & \frac{1}{K} [H(V_X) + H(V_{CX}) + H(V_{CY}) - H(V_{CY}) \nonumber\\
& - & - H(V_{CX})]  - \delta		+ \delta_1	\nonumber\\
& = & \frac{1}{K} H(V_X) -\delta + \delta_1
\label{lemma2_part3}
\end{eqnarray}

where \eqref{lemma2_ref5} holds because $V_Y$ and $X^K$ are independent, \eqref{lemma2_ref6} holds because $V_{CX}$ is a function of $X^K$ and \eqref{lemma2_ref7} holds because $X^K$ is independent of $V_{CY}$ asymptotically and $V_{CX}$ is independent of $V_{CY}$ asymptotically. 

For the proof of \eqref{H_inequality4}, we look at the following probabilities:
\begin{eqnarray}
\text{Pr} \{V_X,V_{CX} \neq G(T_X)\} \le \delta
\label{lemma2_eqn1}
\end{eqnarray}

\begin{eqnarray}
\text{Pr} \{V_Y,V_{CY} \neq G(T_Y)\} \le \delta
\label{lemma2_eqn2}
\end{eqnarray}

\begin{eqnarray}
& & \frac{1}{K} H(X^K|T_Y)											\nonumber\\
& \le & \frac{1}{K} H(X^K, V_{CY},V_Y)] + \delta					\label{lemma2_ref8}\\
& = & \frac{1}{K} [H(X^K, V_{CY},V_{Y}) - H(V_{CY},V_{Y})] + \delta					\nonumber\\
& = & \frac{1}{K} [H(X^K, V_{Y}) - H(V_{CY},V_{Y})] + \delta						\label{lemma2_ref9}\\
& = & \frac{1}{K} [H(X^K|V_{Y}) + I(X^K;V_{Y}) + H(V_{Y}|X^K)] 					\nonumber\\
& & - \frac{1}{K} [H(V_{CY}|V_{Y}) + I(V_{CY};V_{Y}) + H(V_{Y}|V_{CY})]  + \delta		\nonumber\\	
& = & \frac{1}{K} [H(X^K) + H(V_{Y})- H(V_{CY}) - H(V_{Y})]+ \delta	\label{lemma2_ref10}\\
& = & \frac{1}{K} [H(X^K) - H(V_{CY})] + \delta											
\label{lemma2_part4.1}
\end{eqnarray}

where \eqref{lemma2_ref8} holds from \eqref{lemma2_eqn2}, \eqref{lemma2_ref9} holds because $V_{CY}$ and $V_Y$ are asymptotically independent. Furthermore, \eqref{lemma2_ref10} holds because $V_{CY}$ and $V_{Y}$ are asymptotically independent and $X^K$ and $V_{Y}$ are asymptotically independent.

Following a similar proof to those done above in this section, another bound for $H(X^K|V_{CY},V_Y)$ can be found as follows:
\begin{eqnarray}
& & \frac{1}{K} H(X^K|V_{CY},V_Y)											\nonumber\\
& = & \frac{1}{K} [H(X^K,V_{CY},V_{Y}) - H(V_{CY},V_{Y})]					\nonumber\\
& = & \frac{1}{K} [H(X^K,V_{Y}) - H(V_{CY},V_{Y})] 						\label{lemma2_ref11}\\
& = & \frac{1}{K} [H(X^K|V_{Y}) + I(X^K;V_{Y}) + H(V_{Y}|X)] 					\nonumber\\
& & - \frac{1}{K} [H(V_{CY}|V_{Y}) + I(V_{CY};V_{Y}) + H(V_{Y}|V_{CY})] 		\nonumber\\	
& = & \frac{1}{K} [H(X^K) + H(V_{Y})- H(V_{CY}) - H(V_{Y})]				\label{lemma2_ref12}\\
& = & \frac{1}{K} [H(X^K) - H(V_{CY})]	\nonumber\\
& \geq & \frac{1}{K} [H(V_X) + H(V_{CX}) + H(V_{CY}) - H(V_{CY})]	- \delta \nonumber\\
& = & \frac{1}{K} [H(V_X) + H(V_{CX})]	- \delta
\label{lemma2_part4.2}
\end{eqnarray}

where \eqref{lemma2_ref11} and \eqref{lemma2_ref12} hold for the same reason as \eqref{lemma2_ref9} and \eqref{lemma2_ref10} respectively. 

Since we consider the information leakage as the total information obtained subtracted from the total uncertainty, the following hold for the four cases considered in this section:

\begin{eqnarray}
L_{V_X,V_Y}^{X^K} & = & H(X^K) - H(X^K|V_X,V_Y) 	\nonumber\\ 
& \le & H(X^K) - H(V_{CX}) - H(V_{CY}) + \delta
\label{Lemma2_proof_inequality1}
\end{eqnarray}
which proves \eqref{L_inequality1}.

\begin{eqnarray}
L_{V_{CX},V_{CY}}^{X^K} & = & H(X^K) - H(X^K|V_{CX},V_{CY}) \nonumber
\\ & \le & H(X^K) - H(V_X) + \delta
\label{Lemma2_proof_inequality2}
\end{eqnarray}
which proves \eqref{L_inequality2}.

\begin{eqnarray}
L_{V_{CX},V_{CY},V_Y}^{X^K} & = & H(X^K) - H(X^K|V_{CX},V_{CY},V_Y)  \nonumber
\\ & \le & H(X^K) - H(V_X) + \delta
\label{Lemma2_proof_inequality3}
\end{eqnarray}
which proves \eqref{L_inequality3}.

The two bounds for $H(V_{CY},V_Y)$ are given by \eqref{lemma2_part4.1} and \eqref{lemma2_part4.2}. 
From \eqref{lemma2_part4.1}:

\begin{eqnarray}
L_{V_{Y},V_{CY}}^{X^K} & \geq & H(X^K) - [H(X) - H(V_{CY}) + \delta] \nonumber \\
& \geq & H(V_{CY}) - \delta
\label{Lemma2_proof_inequality4.1}
\end{eqnarray}

and from \eqref{lemma2_part4.2}:
\begin{eqnarray}
L_{V_{Y},V_{CY}}^{X^K} & \le & H(X^K) - \left (H(V_X) + H(V_{CX})  - \delta \right)  \nonumber \\
& \le & H(X^K) - H(V_X) - H(V_{CX}) + \delta 
\label{Lemma2_proof_inequality4.2}
\end{eqnarray}

Combining these results from \eqref{Lemma2_proof_inequality4.1} and \eqref{Lemma2_proof_inequality4.2} gives \eqref{L_inequality4}.
\end{proof}

\section{Shannon's Cipher System}

\begin{figure}[ht]
\centering
\includegraphics [scale = 0.7]{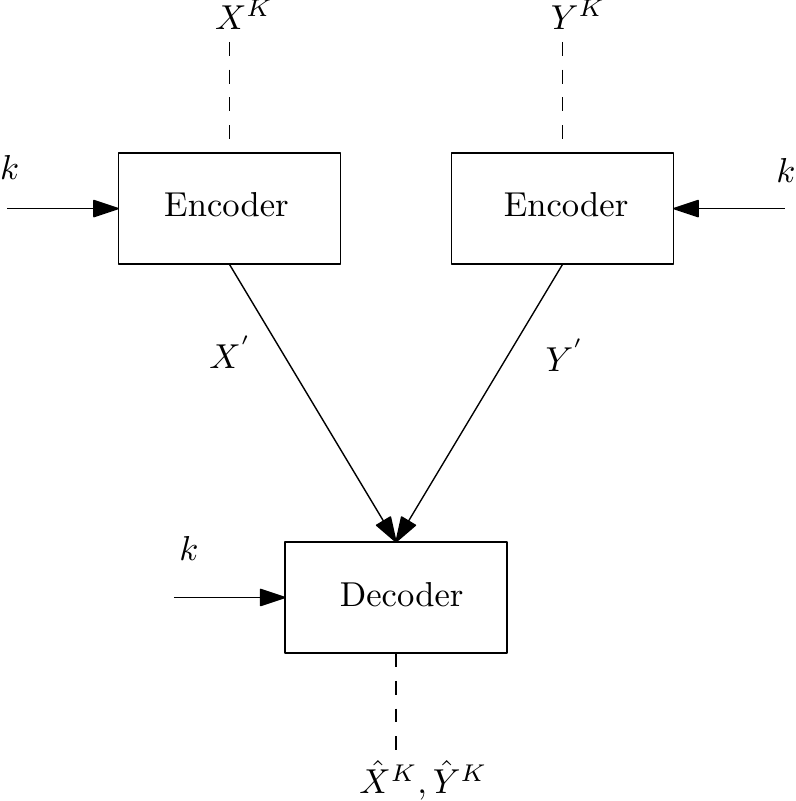}
\caption{Shannon cipher system for two correlated sources}
\label{fig:shannon_cipher_2sources}
\end{figure}

Here, we discuss Shannon's cipher system for two independent correlated sources (depicted in Figure \ref{fig:shannon_cipher_2sources}). The two source outputs are i.i.d random variables $X$ and $Y$, taking on values in the finite sets $\mathcal{X}$ and $\mathcal{Y}$. Both the transmitter and receiver have access to the key, a random variable, independent of $X^K$ and $Y^K$ and taking values in $I_{M_k} = \{0, 1, 2, \ldots ,M_{k} - 1\}$. The sources $X^K$ and $Y^K$ compute the ciphertexts $X^{'}$ and $Y^{'}$, which are the result of specific encryption functions on the plaintext from $X$ and $Y$ respectively. The encryption functions are invertible, thus knowing $X^{'}$ and the key, $X^K$ can be retrieved. 

The mutual information between the plaintext and ciphertext should be small so that the wiretapper cannot gain much information about the plaintext. For perfect secrecy, this mutual information should be zero, then the length of the key should be at least the length of the plaintext.

The encoder functions for $X$ and $Y$, ($E_X$ and $E_Y$ respectively) are given as:

\begin{eqnarray}
E_X : \mathcal{X}^K \times I_{M_{kX}} & \rightarrow & I_{M_X'} =  \{0, 1, \ldots, M_X' - 1\} \nonumber 
\\ && I_{M_{CX}'} =  \{0, 1, \ldots, M_{CX}' - 1\}
\label{xencoder_fcn}
\end{eqnarray}

\begin{eqnarray}
E_Y : \mathcal{Y}^K \times I_{M_{kY}} & \rightarrow & I_{M_Y'} =  \{0, 1, \ldots, M_Y' - 1\} \nonumber 
\\ && I_{M_{CY}'} =  \{0, 1, \ldots, M_{CY}' - 1\}
\label{yencoder_fcn}
\end{eqnarray}

The decoder is defined as:

\begin{eqnarray}
D_{XY} : (I_{M'_X}, I_{M'_Y}, I_{M'_{CX}},I_{M'_{CY}})  & \times &  I_{M_{kX}}, I_{M_{kY}} \nonumber \\
& \rightarrow & \mathcal{X}^K \times \mathcal{Y}^K
\end{eqnarray}

The encoder and decoder mappings are below:
\begin{eqnarray}
W_1 = F_{E_X} (X^K, W_{kX})
\end{eqnarray}

\begin{eqnarray}
W_2 = F_{E_Y} (Y^K, W_{kY})
\end{eqnarray}

\begin{eqnarray}
\widehat{X}^K = F_{D_X} (W_1, W_2, W_{kX})
\end{eqnarray}

\begin{eqnarray}
\widehat{Y}^K = F_{D_Y} (W_1, W_2, W_{kY})
\end{eqnarray}

or 

\begin{eqnarray}
(\widehat{X}^K, \widehat{Y}^K) = F_{D_{XY}} (W_1, W_2, W_{kX}, W_{kY})
\end{eqnarray}

The following conditions should be satisfied for cases 1- 4:

\begin{eqnarray}
\frac{1}{K}\log M_X \le R_X +\epsilon
\label{cond1}
\end{eqnarray}

\begin{eqnarray}
\frac{1}{K}\log M_Y \le R_Y +\epsilon
\label{cond2}
\end{eqnarray}

\begin{eqnarray}
\frac{1}{K}\log M_{kX} \le R_{kX} +\epsilon
\label{cond3}
\end{eqnarray}

\begin{eqnarray}
\frac{1}{K}\log M_{kY} \le R_{{kY}} +\epsilon
\label{cond4}
\end{eqnarray}

\begin{eqnarray}
\text {Pr} \{\widehat{X}^K \neq X^K\} \le \epsilon
\label{cond5}
\end{eqnarray}

\begin{eqnarray}
\text{Pr} \{ \widehat{Y}^K \neq Y^K\} \le \epsilon
\label{cond6}
\end{eqnarray}

\begin{eqnarray}
\frac{1}{K} H(X^K|W_1) \le h_X + \epsilon
\label{cond7}
\end{eqnarray}

\begin{eqnarray}
\frac{1}{K} H(Y^K|W_2) \le h_Y + \epsilon
\label{cond8}
\end{eqnarray}

\begin{eqnarray}
\frac{1}{K} H(X^K,Y^K|W_1) \le h_{XY} + \epsilon
\label{cond8.1}
\end{eqnarray}

\begin{eqnarray}
\frac{1}{K} H(X^K,Y^K|W_2) \le h_{XY} + \epsilon
\label{cond9}
\end{eqnarray}

where $R_X$ is the the rate of source $X$'s channel and $R_Y$ is the the rate of source $Y$'s channel. Here, $R_{kX}$ is the rate of the key channel at $X^K$ and $R_{kY}$ is the rate of the key channel at $Y^K$. The security levels, which are measured by the total and individual uncertainties are $h_{XY}$ and $(h_X, h_Y)$ respectively. 
\\\\
The cases 1 - 5 are:
\\ \textit{Case 1:} When $T_X$ and $T_Y$ are leaked and both $X^K$ and $Y^K$ need to be kept secret.
\\ \textit{Case 2:} When $T_X$ and $T_Y$ are leaked and $X^K$ needs to be kept secret.
\\ \textit{Case 3:} When $T_X$ is leaked and both $X^K$ and $Y^K$ need to be kept secret.
\\ \textit{Case 4:} When $T_X$ is leaked and $Y^K$ needs to be kept secret.
\\ \textit{Case 5:} When $T_X$ is leaked and $X^K$ needs to be kept secret.
\\ where $T_X$ is the syndrome produced by $X$, containing $V_{CX}$ and $V_X$ and $T_Y$ is the syndrome produced by $Y$, containing $V_{CY}$ and $V_X$ .
\\\\
The admissible rate region for each case is defined as follows:
\\ \textit{Definition 1a:} ($R_X$, $R_Y$, $R_{kX}$, $R_{kY}$, $h_{XY}$) is admissible for case 1 if there exists a code ($F_{E_{X}}$, $F_{D_{XY}}$) and ($F_{E_{Y}}$, $F_{D_{XY}}$) such that \eqref{cond1} - \eqref{cond6} and \eqref{cond9} hold for any $\epsilon \rightarrow 0$ and sufficiently large $K$.
\\ \textit{Definition 1b:} ($R_X$, $R_Y$, $R_{kX}$, $R_{kY}$, $h_{X}$) is admissible for case 2 if there exists a code ($F_{E_{X}}$, $F_{D_{XY}}$) such that \eqref{cond1} - \eqref{cond7} hold for any $\epsilon \rightarrow 0$ and sufficiently large $K$.
\\ \textit{Definition 1c:} ($R_X$, $R_Y$, $R_{kX}$, $R_{kY}$, $h_{X}$, $h_{Y}$) is admissible for case 3 if there exists a code ($F_{E_{X}}$, $F_{D_{XY}}$) and ($F_{E_{Y}}$, $F_{D_{XY}}$) such that \eqref{cond1} - \eqref{cond6} and \eqref{cond8}, \eqref{cond9} hold for any $\epsilon \rightarrow 0$ and sufficiently large $K$.
\\ \textit{Definition 1d:} ($R_X$, $R_Y$, $R_{kX}$, $R_{kY}$, $h_{Y}$) is admissible for case 4 if there exists a code ($F_{E_{X}}$, $F_{D_{XY}}$) such that \eqref{cond1} - \eqref{cond6} and \eqref{cond8} hold for any $\epsilon \rightarrow 0$ and sufficiently large $K$.
\\ \textit{Definition 1e:} ($R_X$, $R_Y$, $R_{kX}$, $R_{kY}$, $h_{X}$) is admissible for case 5 if there exists a code ($F_{E_{X}}$, $F_{D_{XY}}$) such that \eqref{cond1} - \eqref{cond6} and \eqref{cond7} hold for any $\epsilon \rightarrow 0$ and sufficiently large $K$.
\\ \textit{Definition 2:} The admissible rate regions of $\mathcal{R}_j$ and of $\mathcal{R}_k$ are defined as:

\begin{eqnarray}
\mathcal{R}_1(h_{XY}) = \{(R_X, R_Y, R_{kX}, R_{kY}):			\nonumber
\\(R_X, R_Y, R_{kX}, R_{kY}, h_{XY} ) \text{ is admissible for case 1} \}
\end{eqnarray}

\begin{eqnarray}
\mathcal{R}_2(h_{X}) = \{(R_X, R_Y, R_{kX}, R_{kY}):			\nonumber
\\ (R_X, R_Y, R_{kX}, R_{kY}, h_{X} ) \text{ is admissible for case 2} \}
\end{eqnarray}

\begin{eqnarray}
\mathcal{R}_3(h_X, h_Y) = \{(R_X, R_Y, R_{kX}, R_{kY}):			\nonumber
\\ (R_X, R_Y, R_{kX}, R_{kY}, h_{X}, h_{Y} ) \text{ is admissible for case 3} \}
\end{eqnarray}

\begin{eqnarray}
\mathcal{R}_4(h_{Y}) = \{(R_X, R_Y, R_{kX}, R_{kY}):			\nonumber
\\(R_X, R_Y, R_{kX}, R_{kY}, h_{Y} ) \text{ is admissible for case 4} \}
\end{eqnarray}

\begin{eqnarray}
\mathcal{R}_5(h_{X}) = \{(R_X, R_Y, R_{kX}, R_{kY}):			\nonumber
\\(R_X, R_Y, R_{kX}, R_{kY}, h_{X} ) \text{ is admissible for case 5} \}
\end{eqnarray}

Theorems for these regions have been developed:

\textit{Theorem 2:} For $0 \le h_{XY} \le H(X,Y)$,
\begin{eqnarray}
&& \mathcal{R}_1(h_{XY}) = \{(R_X, R_Y, R_{kX},R_{kY}): 		\nonumber
\\ && R_X \geq H(X|Y), 				\nonumber
\\ && R_Y \geq H(Y|X),		\nonumber
\\ && R_X + R_Y	\geq H(X,Y)						\nonumber
\\ && R_{kX} \geq h_{XY} \text{ and } R_{kY} \geq h_{XY} \}			
\label{theorem2}
\end{eqnarray}

\textit{Theorem 3:} For $0 \le h_{X} \le H(X)$,
\begin{eqnarray}
&& \mathcal{R}_2(h_{X}) = \{(R_X, R_Y, R_{kX},R_{kY}): 		\nonumber
\\ && R_X \geq H(X|Y), 				\nonumber
\\ && R_Y \geq H(Y|X),		\nonumber
\\ && R_X + R_Y	\geq H(X,Y)						\nonumber
\\ && R_{kX} \geq h_X \text{ and } R_{kY} \geq h_Y \}			
\label{theorem3}
\end{eqnarray}

\textit{Theorem 4:} For $0 \le h_{X} \le H(X)$ and $0 \le h_{Y} \le H(Y)$,
\begin{eqnarray}
&& \mathcal{R}_3(h_{X}, h_{Y}) = \{(R_X, R_Y, R_{kX},R_{kY}): 		\nonumber
\\ && R_X \geq H(X|Y), 				\nonumber
\\ && R_Y \geq H(Y|X),		\nonumber
\\ && R_X + R_Y	\geq H(X,Y)						\nonumber
\\ && R_{kX} \geq h_{X} \text{ and } R_{kY} \geq h_{Y} \}
\label{theorem4}
\end{eqnarray}

\textit{Theorem 5:} For $0 \le h_{X} \le H(X)$,
\begin{eqnarray}
&& \mathcal{R}_5(h_{X}, h_{Y}) = \{(R_X, R_Y, R_{kX},R_{kY}): 		\nonumber
\\ && R_X \geq H(X|Y), 				\nonumber
\\ && R_Y \geq H(Y|X),		\nonumber
\\ && R_X + R_Y	\geq H(X,Y)						\nonumber
\\ && R_{kX} \geq h_{X} \text{ and } R_{kY} \geq 0 \}
\label{theorem5}
\end{eqnarray}

When $h_X = 0$ then case $5$ can be reduced to that depicted in \eqref{theorem4}.  
Hence, Corollary 1 follows:
\\ \textit{Corollary 1:} $\mathcal{R}_4(h_{Y}) = \mathcal{R}_3(0, h_Y)$

The security levels, which are measured by the total and individual uncertainties $h_{XY}$ and $(h_X, h_Y)$ respectively give an indication of the level of uncertainty in knowing certain information. When the uncertainty increases then less information is known to an eavesdropper and there is a higher level of security.

\section{Proof of Theorems 2 - 5}
This section initially proves the direct parts of Theorems 2 - 5 and thereafter the converse parts.

\subsection{Direct parts}
All the channel rates in the theorems above are in accordance with Slepian-Wolf's theorem, hence there is no need to prove them. 
We construct a code based on the prototype code ($W_X, W_Y, W_{CX}, W_{CY}$) in Lemma 1. In order to include a key in the prototype code, $W_X$ is divided into two parts as per the method used by Yamamoto \cite{shannon1_yamamoto}:
\begin{eqnarray}
W_{X1} = W_X \text{ mod } M_{X1} \in I_{M_{X1}} = \{0, 1, 2, \ldots, M_{X1} - 1\}
\label{theorems2-4_eq_1}
\end{eqnarray}

\begin{eqnarray}
W_{X2} = \frac{W_X - W_{X1}}{M_{X1}} \in I_{M_{X2}} = \{0, 1, 2, \ldots, M_{X2} - 1\}
\label{theorems2-4_eq_2}
\end{eqnarray}

where $M_{X1}$ is a given integer and $M_{X2}$ is the ceiling of $M_X/M_{X1}$. The $M_X/M_{X1}$ is considered an integer for simplicity, because the difference between the ceiling value and the actual value can be ignored when $K$ is sufficiently large. In the same way, $W_Y$ is divided:

\begin{eqnarray}
W_{Y1} = W_Y \text{ mod } M_{Y1} \in I_{M_{Y1}} = \{0, 1, 2, \ldots, M_{Y1} - 1\}
\label{theorems2-4_eq_3}
\end{eqnarray}

\begin{eqnarray}
W_{Y2} = \frac{W_Y - W_{Y1}}{M_{Y1}} \in I_{M_{Y2}} = \{0, 1, 2, \ldots, M_{Y2} - 1\}
\label{theorems2-4_eq_4}
\end{eqnarray}

The common information components $W_{CX}$ and $W_{CY}$ are already portions and are not divided further. 
It can be shown that when some of the codewords are wiretapped the uncertainties of $X^K$ and $Y^K$ are bounded as follows:

\begin{eqnarray}
\frac{1}{K} H(X^K|W_{X2},W_Y) \geq I(X;Y) + \frac{1}{K} \log M_{X1} - \epsilon_{0}^{'}
\label{theorems2-4_ineq_1}
\end{eqnarray}

\begin{eqnarray}
\frac{1}{K} H(Y^K|W_{X},W_{Y2}) \geq I(X;Y) + \frac{1}{K} \log M_{Y1} - \epsilon_{0}^{'}
\label{theorems2-4_ineq_2}
\end{eqnarray}

\begin{eqnarray}
\frac{1}{K} H(X^K|W_{X},W_{Y2}) \geq I(X;Y) - \epsilon_{0}^{'}
\label{theorems2-4_ineq_3}
\end{eqnarray}

\begin{eqnarray}
\frac{1}{K} H(X^K|W_{X},W_Y, W_{CY}) \geq \frac{1}{K} \log M_{CX} - \epsilon_{0}^{'}
\label{theorems2-4_ineq_4}
\end{eqnarray}

\begin{eqnarray}
\frac{1}{K} H(Y^K|W_{X},W_Y, W_{CY}) \geq \frac{1}{K} \log M_{CX} - \epsilon_{0}^{'}
\label{theorems2-4_ineq_5}
\end{eqnarray}

\begin{eqnarray}
\frac{1}{K} H(X^K|W_Y, W_{CY}) \geq H(X|Y) + \frac{1}{K} \log M_{CX} - \epsilon_{0}^{'}
\label{theorems2-4_ineq_6}
\end{eqnarray}

\begin{eqnarray}
\frac{1}{K} H(Y^K|W_Y, W_{CY}) \geq \frac{1}{K} \log M_{CX} - \epsilon_{0}^{'}
\label{theorems2-4_ineq_7}
\end{eqnarray}

where $\epsilon_{0}^{'} \rightarrow 0$ as  $\epsilon_{0} \rightarrow 0$.
The proofs for \eqref{theorems2-4_ineq_1} - \eqref{theorems2-4_ineq_7} are the same as per Yamamoto's\cite{shannon1_yamamoto} proof in Lemma A1. The difference is that $W_{CX}$, $W_{CY}$, $M_{CX}$ and $M_{CY}$ are described as $W_{C1}$, $W_{C2}$, $M_{C1}$ and $M_{C2}$ respectively by Yamamoto. Here, we consider that $W_{CX}$ and $W_{CY}$ are represented by Yamamoto's $W_{C1}$ and $W_{C2}$ respectively. In addition there are some more inequalities considered here:
\begin{eqnarray}
 \frac{1}{K} H(Y^K|W_X, W_{CX}, W_{CY}, W_{Y2}) & \geq & \frac{1}{K} \log M_{Y1}  \nonumber
 \\ & - & \epsilon_{0}^{'}
\label{theorems2-4_ineq_8}
\end{eqnarray}

\begin{eqnarray}
 \frac{1}{K} H(Y^K|W_X, W_{CX}, W_{CY}) & \geq & \frac{1}{K} \log M_{Y1}  \nonumber
\\ & + & \frac{1}{K} \log M_{Y2} - \epsilon_{0}^{'}
\label{theorems2-4_ineq_9}
\end{eqnarray}

\begin{eqnarray}
\frac{1}{K} H(X^K|W_{X2}, W_{CY}) & \geq & \frac{1}{K} \log M_{X1} 	\nonumber
\\ & + & \frac{1}{K} \log M_{CX} - \epsilon_{0}^{'}
\label{theorems2-4_ineq_10}
\end{eqnarray}

\begin{eqnarray}
\frac{1}{K} H(Y^K|W_{X2}, W_{CY}) & \geq & \frac{1}{K} \log M_{Y1} 	\nonumber
\\ & + & \frac{1}{K} \log M_{Y2} + \frac{1}{K} \log M_{CX} 			\nonumber
\\ & - & \epsilon_{0}^{'}
\label{theorems2-4_ineq_11}
\end{eqnarray}

The inequalities \eqref{theorems2-4_ineq_8} and \eqref{theorems2-4_ineq_9} can be proved in the same way as per Yamamoto's\cite{shannon1_yamamoto} Lemma A2, and  \eqref{theorems2-4_ineq_10} and \eqref{theorems2-4_ineq_11} can be proved in the same way as per Yamamoto's\cite{shannon1_yamamoto} Lemma A1. 

For each proof we consider cases where a key already exists for either $V_{CX}$ or $V_{CY}$ and the encrypted common information portion is then used to mask the other portions (either $V_{CX}$ or $V_{CY}$ and the private information portions). There are two cases considered for each; firstly, when the common information portion entropy is greater than the entropy of the portion that needs to be masked, and secondly when the common information portion entropy is less than the entropy of the portion to be masked. For the latter case, a smaller key will need to be added so as to cover the portion entirely. This has the effect of reducing the required key length, which is explained in greater detail in Section VII.

\begin{proof}[Proof of Theorem 2]
Suppose that ($R_X$, $R_Y$, $R_{KX}$, $R_{KY}$) $\in$ 
$\mathcal{R}_1$ for $h_{XY} \le H(X,Y)$. Without loss of generality, we assume that $h_X \le h_Y$. Then, from \eqref{theorem2} 
\begin{eqnarray}
&& R_X \geq H(X^K|Y^K)  				\nonumber
\\&& R_Y  \geq H(Y^K|X^K) 				\nonumber
\\&& R_X + R_Y  \geq H(X^K, Y^K)
\label{theorem2_proof_1}
\end{eqnarray}

\begin{eqnarray}
R_{kX} \geq h_{XY}, R_{kY} \geq h_{XY} 
\label{theorem2_proof_2}
\end{eqnarray}

Assuming a key exists for $V_{CY}$. For the first case, consider the following: $H(V_{CY}) \geq H(V_X)$, $H(V_{CY}) \geq H(V_Y)$ and $H(V_{CY}) \geq H(V_{CX})$.

\begin{eqnarray}
M_{CY} = 2^{K h_{XY}}
\label{theorem2_proof_6}
\end{eqnarray}

The codewords $W_X$ and $W_Y$ and their keys $W_{kX}$ and $W_{kY}$ are now defined:

\begin{eqnarray}
W_X = (W_{X1} \oplus W_{kCY}, W_{X2}  \oplus W_{kCY}, W_{CX}  \oplus W_{kCY})
\label{theorem2_proof_7}
\end{eqnarray}

\begin{eqnarray}
W_Y = (W_{Y1} \oplus W_{kCY}, W_{Y2}  \oplus W_{kCY}, W_{CY})
\label{theorem2_proof_8}
\end{eqnarray}

\begin{eqnarray}
W_{kY} = (W_{kCY})
\label{theorem2_proof_10}
\end{eqnarray}

where $W_\alpha \in I_{M_\alpha} = \{0, 1, \ldots, M_\alpha - 1\}$. The wiretapper will not know $W_{X1}$, $W_{X2}$ and $W_{CX}$ from $W_X$ and $W_{Y1}$, $W_{Y2}$ and $W_{CY}$ from $W_Y$ as these are protected by the key ($W_{kCY}$.

In this case, $R_X$, $R_Y$, $R_{kX}$ and $R_{kY}$ satisfy from \eqref{lemma1_2} - \eqref{lemma1_4} and \eqref{theorem2_proof_1} - \eqref{theorem2_proof_6}, that

\begin{eqnarray}
\frac{1}{K} \log M_X + \frac{1}{K} \log M_Y  & = & \frac{1}{K} (\log M_{X1} + \log M_{X2}  \nonumber
\\ & + &  \log M_{CX}) + \frac{1}{K} (\log M_{Y1}  \nonumber
\\ & + &  \log M_{Y2} +  \log M_{CY}) \nonumber 
\\ & \le & H(X|Y) + H(Y|X) 				\nonumber
\\ & + & I(X;Y) + \epsilon_0 \nonumber
\\ & = & H(X,Y)  		\nonumber
\\ & \le & R_X + R_Y
\label{theorem2_proof_11}
\end{eqnarray}

\begin{eqnarray}
\frac{1}{K} \log M_{kX} & = & \frac{1}{K} \log M_{CY}	\nonumber
\\ & = & h_{XY}			\label{num3}
\\ & \le & R_{kX}
\label{theorem2_proof_13}
\end{eqnarray}

where \eqref{num3} comes from \eqref{theorem2_proof_6}.

\begin{eqnarray}
\frac{1}{K} \log M_{kY} & = & \frac{1}{K} \log M_{CY}	\nonumber
\\ & = & h_{XY}			\label{num4}
\\ & \le & R_{kY}
\label{theorem2_proof_14}
\end{eqnarray}
where \eqref{num4} comes from \eqref{theorem2_proof_6}.

The security levels thus result:
\begin{eqnarray}
\frac{1}{K} H(X^K|W_X, W_Y) & = & \frac{1}{K} H(X|W_{X1} \oplus W_{kCY}, \nonumber
\\ && W_{X2} \oplus W_{kCY}, W_{CX} \oplus W_{kCY},			\nonumber
\\ && W_{Y1} \oplus W_{kCY}, W_{Y2} \oplus W_{kCY},	\nonumber
\\ &&  W_{CY}) 				\nonumber
\\ & = & H(X^K)	\label{num5}
\\ & \ge & h_X - \epsilon^{'}_0
\label{theorem2_proof_16}
\end{eqnarray}

where \eqref{num5} holds because $W_{X1}$, $W_{X2}$, $W_{CX}$, $W_{Y1}$, $W_{Y2}$ are covered by key $W_{CY}$ and $W_{CY}$ is covered by an existing random number key. Equations \eqref{lemma1_1} - \eqref{lemma1_7} imply that $W_{X1}$, $W_{X2}$, $W_{Y1}$ and $W_{Y2}$ have almost no redundancy and they are mutually independent.

Similarly, 
\begin{eqnarray}
\frac{1}{K} H(Y^K|W_X, W_Y) \geq h_Y - \epsilon^{'}_0
\label{theorem2_proof_17}
\end{eqnarray}

Therefore ($R_X$, $R_Y$, $R_{kX}$, $R_{kY}$, $h_{XY}$, $h_{XY}$) is admissible from \eqref{theorem2_proof_11} -  \eqref{theorem2_proof_17}.

Next the case where: $H(V_{CY}) < H(V_X)$, $H(V_{CY}) < H(V_Y)$ and $H(V_{CY}) < H(V_{CX})$ is considered. Here, there are shorter length keys used in addition to the key provided by $W_{CY}$ in order to make the key lengths required by the individual portions. For example the key $W_{k1}$ comprises $W_{kCY}$ and a short key $W_1$, which together provide the length of $W_{X1}$.
The codewords $W_X$ and $W_Y$ and their keys $W_{kX}$ and $W_{kY}$ are now defined:

\begin{eqnarray}
W_X = (W_{X1} \oplus W_{k1}, W_{X2}  \oplus W_{k2}, W_{CX}  \oplus W_{k3})
\label{theorem2_proof_7.1.1}
\end{eqnarray}

\begin{eqnarray}
W_Y = (W_{Y1} \oplus W_{k4}, W_{Y2}  \oplus W_{k5}, W_{CY})
\label{theorem2_proof_8.1.1}
\end{eqnarray}

\begin{eqnarray}
W_{kX} = (W_{k1}, W_{k2}, W_{k3})
\label{theorem2_proof_10.1.1}
\end{eqnarray}

\begin{eqnarray}
W_{kY} = (W_{k4}, W_{k5})
\label{theorem2_proof_10.1.2}
\end{eqnarray}

where $W_\alpha \in I_{M_\alpha} = \{0, 1, \ldots, M_\alpha - 1\}$. The wiretapper will not know $W_{X1}$, $W_{X2}$ and $W_{CX}$ from $W_X$ and $W_{Y1}$, $W_{Y2}$ and $W_{CY}$ from $W_Y$ as these are protected by the key ($W_{kCY}$.

In this case, $R_X$, $R_Y$, $R_{kX}$ and $R_{kY}$ satisfy that

\begin{eqnarray}
\frac{1}{K} \log M_X + \frac{1}{K} \log M_Y  & = & \frac{1}{K} (\log M_{X1} + \log M_{X2}  \nonumber
\\ & + &  \log M_{CX}) + \frac{1}{K} (\log M_{Y1}  \nonumber
\\ & + &  \log M_{Y2} +  \log M_{CY}) \nonumber 
\\ & \le & H(X|Y) + H(Y|X) 				\nonumber
\\ & + & I(X;Y) + \epsilon_0 \nonumber
\\ & = & H(X,Y)  		\nonumber
\\ & \le & R_X + R_Y
\label{theorem2_proof_11.1.1}
\end{eqnarray}

\begin{eqnarray}
\frac{1}{K} \log M_{kX} & = & \frac{1}{K} [\log M_{k1} + \log M_{k2} + \log M_{k3}]	\nonumber
\\ & = & \log M_{kCY} + \log M_{1} \nonumber
\\ & + & \log M_{kCY} + \log M_{2} \nonumber
\\ & + & \log M_{kCY} + \log M_{3} \nonumber
\\ & = & 3 \log M_{kCY} + \log M_{1} \nonumber
\\ & + & \log M_{2} + \log M_{3} \nonumber
\\ & \geq & 3 h_{XY} - \epsilon_0  \label{num333.1}
\\ & \geq & h_{XY}
\label{theorem2_proof_13.1.1}
\end{eqnarray}

where \eqref{num333.1} results from \eqref{theorem2_proof_6}.

\begin{eqnarray}
\frac{1}{K} \log M_{kX} & = & \frac{1}{K} [\log M_{k3} + \log M_{k4} + \log M_{kCY}]	\nonumber
\\ & = & \log M_{kCY} + \log M_{3} +	\log M_{kCY} \nonumber
\\ & + & \log M_{4} + \log M_{kCY} 		\nonumber
\\ & = & 3 \log M_{kCY} + \log M_{3} + \log M_{4}
\\ & \geq & 3 h_{XY} - \epsilon_0 \label{num333}
\\ & \geq & h_{XY}
\label{theorem2_proof_14.1.1}
\end{eqnarray}

where \eqref{num333} results from \eqref{theorem2_proof_6}.

The security levels thus result:
\begin{eqnarray}
\frac{1}{K} H(X^K|W_X, W_Y) & = & \frac{1}{K} H(X|W_{X1} \oplus W_{k1}, W_{X2} \oplus W_{k2}, \nonumber
\\ && W_{CX} \oplus W_{k3},			\nonumber
\\ && W_{Y1} \oplus W_{k4}, W_{Y2} \oplus W_{k5},	\nonumber
\\ &&  W_{CY}) 				\nonumber
\\ & = & H(X^K)	\label{num5.1.1}
\\ & \ge & h_X - \epsilon^{'}_0
\label{theorem2_proof_16.1.1}
\end{eqnarray}

where \eqref{num5} holds because $W_{X1}$, $W_{X2}$, $W_{CX}$, $W_{Y1}$, $W_{Y2}$ are covered by key $W_{CY}$ and some shorter length key and $W_{CY}$ is covered by an existing random number key. 

Similarly, 
\begin{eqnarray}
\frac{1}{K} H(Y^K|W_X, W_Y) \geq h_Y - \epsilon^{'}_0
\label{theorem2_proof_17.1.1}
\end{eqnarray}

Therefore ($R_X$, $R_Y$, $R_{kX}$, $R_{kY}$, $h_{XY}$, $h_{XY}$) is admissible from \eqref{theorem2_proof_11.1.1} -  \eqref{theorem2_proof_17.1.1}.

\end{proof}

Theorem 3 - 5 are proven in the same way with varying codewords and keys. The proofs follow:

\begin{proof}[Theorem 3 proof]

The consideration for the security levels is that $h_Y \geq h_X$ because $Y$ contains the key the is used for masking.
Suppose that ($R_X$, $R_Y$, $R_{KX}$, $R_{KY}$) $\in$ 
$\mathcal{R}_2$. From \eqref{theorem3} 
\begin{eqnarray}
&& R_X \geq H(X^K|Y^K)  				\nonumber
\\&& R_Y  \geq H(Y^K|X^K) 				\nonumber
\\&& R_X + R_Y  \geq H(X^K, Y^K)
\label{theorem3_proof_1}
\end{eqnarray}

\begin{eqnarray}
R_{kX} \geq h_{X}, R_{kY} \geq h_{Y} 
\label{theorem3_proof_2}
\end{eqnarray}

Assuming a key exists for $V_{CY}$. For the first case, consider the following: $H(V_{CY}) \geq H(V_X)$, $H(V_{CY}) \geq H(V_Y)$ and $H(V_{CY}) \geq H(V_{CX})$.

\begin{eqnarray}
M_{CY} = 2^{K h_{Y}}
\label{theorem3_proof_6}
\end{eqnarray}

The codewords $W_X$ and $W_Y$ and their keys $W_{kX}$ and $W_{kY}$ are now defined:

\begin{eqnarray}
W_X = (W_{X1} \oplus W_{kCY}, W_{X2}  \oplus W_{kCY}, W_{CX}  \oplus W_{kCY})
\label{theorem3_proof_7}
\end{eqnarray}

\begin{eqnarray}
W_Y = (W_{Y1} \oplus W_{kCY}, W_{Y2}  \oplus W_{kCY}, W_{CY})
\label{theorem3_proof_8}
\end{eqnarray}

\begin{eqnarray}
W_{kY} = (W_{kCY})
\label{theorem3_proof_10}
\end{eqnarray}

where $W_\alpha \in I_{M_\alpha} = \{0, 1, \ldots, M_\alpha - 1\}$. The wiretapper will not know $W_{X1}$, $W_{X2}$ and $W_{CX}$ from $W_X$ and $W_{Y1}$, $W_{Y2}$ and $W_{CY}$ from $W_Y$ as these are protected by the key ($W_{kCY}$.

In this case, $R_X$, $R_Y$, $R_{kX}$ and $R_{kY}$ satisfy from \eqref{lemma1_2} - \eqref{lemma1_4} and \eqref{theorem3_proof_1} - \eqref{theorem3_proof_6}, that

\begin{eqnarray}
\frac{1}{K} \log M_X + \frac{1}{K} \log M_Y  & = & \frac{1}{K} (\log M_{X1} + \log M_{X2}  \nonumber
\\ & + &  \log M_{CX}) + \frac{1}{K} (\log M_{Y1}  \nonumber
\\ & + &  \log M_{Y2} +  \log M_{CY}) \nonumber 
\\ & \le & H(X|Y) + H(Y|X) 				\nonumber
\\ & + & I(X;Y) + \epsilon_0 \nonumber
\\ & = & H(X,Y)  		\nonumber
\\ & \le & R_X + R_Y
\label{theorem3_proof_11}
\end{eqnarray}

\begin{eqnarray}
\frac{1}{K} \log M_{kX} & = & \frac{1}{K} \log M_{CY}	\nonumber
\\ & = & h_{Y}			\label{num3.2}
\\ & \geq & h_X - \epsilon_0  \label{num3.3}
\\ &  & R_{kX}
\label{theorem3_proof_13}
\end{eqnarray}

where \eqref{num3.2} comes from \eqref{theorem3_proof_6} and \eqref{num3.3} comes form the consideration stated at the beginning of this proof.

\begin{eqnarray}
\frac{1}{K} \log M_{kY} & = & \frac{1}{K} \log M_{CY}	\nonumber
\\ & = & h_{XY}			\label{num4.1}
\\ & \le & R_{kY}
\label{theorem3_proof_14}
\end{eqnarray}
where \eqref{num4.1} comes from \eqref{theorem3_proof_6}.

The security levels thus result:
\begin{eqnarray}
\frac{1}{K} H(X^K|W_X, W_Y) & = & \frac{1}{K} H(X|W_{X1} \oplus W_{kCY}, \nonumber
\\ && W_{X2} \oplus W_{kCY}, W_{CX} \oplus W_{kCY},			\nonumber
\\ && W_{Y1} \oplus W_{kCY}, W_{Y2} \oplus W_{kCY},	\nonumber
\\ &&  W_{CY}) 				\nonumber
\\ & = & H(X^K)	\label{num5.1}
\\ & \ge & h_X - \epsilon^{'}_0
\label{theorem3_proof_16}
\end{eqnarray}

where \eqref{num5.1} holds because $W_{X1}$, $W_{X2}$, $W_{CX}$, $W_{Y1}$, $W_{Y2}$ are covered by key $W_{CY}$ and $W_{CY}$ is covered by an existing random number key. Equations \eqref{lemma1_1} - \eqref{lemma1_7} imply that $W_{X1}$, $W_{X2}$, $W_{Y1}$ and $W_{Y2}$ have almost no redundancy and they are mutually independent.

Similarly, 
\begin{eqnarray}
\frac{1}{K} H(Y^K|W_X, W_Y) \geq h_Y - \epsilon^{'}_0
\label{theorem3_proof_17}
\end{eqnarray}

Therefore ($R_X$, $R_Y$, $R_{kX}$, $R_{kY}$, $h_{XY}$, $h_{XY}$) is admissible from \eqref{theorem2_proof_11} -  \eqref{theorem2_proof_17}.

Next the case where: $H(V_{CY}) < H(V_X)$, $H(V_{CY}) < H(V_Y)$ and $H(V_{CY}) < H(V_{CX})$ is considered. Here, there are shorter length keys used in addition to the key provided by $W_{CY}$ in order to make the key lengths required by the individual portions. For example the key $W_{k1}$ comprises $W_{kCY}$ and a short key $W_1$, which together provide the length of $W_{X1}$.
The codewords $W_X$ and $W_Y$ and their keys $W_{kX}$ and $W_{kY}$ are now defined:

\begin{eqnarray}
W_X = (W_{X1} \oplus W_{k1}, W_{X2}  \oplus W_{k2}, W_{CX}  \oplus W_{k3})
\label{theorem3_proof_7.1.1}
\end{eqnarray}

\begin{eqnarray}
W_Y = (W_{Y1} \oplus W_{k4}, W_{Y2}  \oplus W_{k5}, W_{CY})
\label{theorem3_proof_8.1.1}
\end{eqnarray}

\begin{eqnarray}
W_{kX} = (W_{k1}, W_{k2}, W_{k3})
\label{theorem3_proof_10.1.1}
\end{eqnarray}

\begin{eqnarray}
W_{kY} = (W_{k4}, W_{k5})
\label{theorem3_proof_10.1.2}
\end{eqnarray}

where $W_\alpha \in I_{M_\alpha} = \{0, 1, \ldots, M_\alpha - 1\}$. The wiretapper will not know $W_{X1}$, $W_{X2}$ and $W_{CX}$ from $W_X$ and $W_{Y1}$, $W_{Y2}$ and $W_{CY}$ from $W_Y$ as these are protected by the key ($W_{kCY}$.

In this case, $R_X$, $R_Y$, $R_{kX}$ and $R_{kY}$ satisfy that

\begin{eqnarray}
\frac{1}{K} \log M_X + \frac{1}{K} \log M_Y  & = & \frac{1}{K} (\log M_{X1} + \log M_{X2}  \nonumber
\\ & + &  \log M_{CX}) + \frac{1}{K} (\log M_{Y1}  \nonumber
\\ & + &  \log M_{Y2} +  \log M_{CY}) \nonumber 
\\ & \le & H(X|Y) + H(Y|X) 				\nonumber
\\ & + & I(X;Y) + \epsilon_0 \nonumber
\\ & = & H(X,Y)  		\nonumber
\\ & \le & R_X + R_Y
\label{theorem3_proof_11.1.1}
\end{eqnarray}

\begin{eqnarray}
\frac{1}{K} \log M_{kX} & = & \frac{1}{K} [\log M_{k1} + \log M_{k2} + \log M_{k3}]	\nonumber
\\ & = & \log M_{kCY} + \log M_{1} +	\log M_{kCY}  \nonumber
\\ & + &  \log M_{2} + \log M_{kCY} + \log M_{3} \nonumber
\\ & = & 3 \log M_{kCY} + \log M_{1} + \log M_{2} + \log M_{3} \nonumber
\\ & \geq & 3 h_{Y} - \epsilon_0  \label{num334.1}
\\ & \geq & 3 h_{X} - \epsilon_0 \nonumber
\\ & \geq & h_{X}
\label{theorem3_proof_13.1.1}
\end{eqnarray}

where \eqref{num334.1} results from \eqref{theorem3_proof_6} and the result is from the consideration at the beginning of this proof.

\begin{eqnarray}
\frac{1}{K} \log M_{kY} & = & \frac{1}{K} [\log M_{k3} + \log M_{k4} + \log M_{kCY}]	\nonumber
\\ & = & \log M_{kCY} + \log M_{3} +	\log M_{kCY}  \nonumber
\\ & + & \log M_{4} + \log M_{kCY}  \nonumber
\\ & = & 3 \log M_{kCY} + \log M_{3} + \log M_{4} \nonumber
\\ & \geq & 3 h_{Y} - \epsilon_0 \label{num333.5}
\\ & \geq & h_{Y}
\label{theorem3_proof_14.1.1}
\end{eqnarray}

where \eqref{num333.5} results from \eqref{theorem3_proof_6}.

The security levels thus result:
\begin{eqnarray}
\frac{1}{K} H(X^K|W_X, W_Y) & = & \frac{1}{K} H(X|W_{X1} \oplus W_{k1}, \nonumber 
\\ && W_{X2} \oplus W_{k2}, W_{CX} \oplus W_{k3},			\nonumber
\\ && W_{Y1} \oplus W_{k4}, W_{Y2} \oplus W_{k5},	\nonumber
\\ &&  W_{CY}) 				\nonumber
\\ & = & H(X^K)	\label{num5.1.1.1}
\\ & \ge & h_X - \epsilon^{'}_0
\label{theorem3_proof_16.1.1}
\end{eqnarray}

where \eqref{num5.1.1.1} holds because $W_{X1}$, $W_{X2}$, $W_{CX}$, $W_{Y1}$, $W_{Y2}$ are covered by key $W_{CY}$ and some shorter length key and $W_{CY}$ is covered by an existing random number key. 

Similarly, 
\begin{eqnarray}
\frac{1}{K} H(Y^K|W_X, W_Y) \geq h_Y - \epsilon^{'}_0
\label{theorem3_proof_17.1.1}
\end{eqnarray}

Therefore ($R_X$, $R_Y$, $R_{kX}$, $R_{kY}$, $h_{XY}$, $h_{XY}$) is admissible from \eqref{theorem2_proof_11.1.1} -  \eqref{theorem2_proof_17.1.1}.

\end{proof}

\begin{proof}[Proof of Theorem 4]

Again, the consideration for the security levels is that $h_Y \geq h_X$ because $Y$ contains the key the is used for masking.
Suppose that ($R_X$, $R_Y$, $R_{KX}$, $R_{KY}$) $\in$ 
$\mathcal{R}_3$. From \eqref{theorem3} 
\begin{eqnarray}
&& R_X \geq H(X^K|Y^K)  				\nonumber
\\&& R_Y  \geq H(Y^K|X^K) 				\nonumber
\\&& R_X + R_Y  \geq H(X^K, Y^K)
\label{theorem4_proof_1}
\end{eqnarray}

\begin{eqnarray}
R_{kX} \geq h_{X}, R_{kY} \geq h_{Y} 
\label{theorem4_proof_2}
\end{eqnarray}

Assuming a key exists for $V_{CY}$. For the first case, consider the following: $H(V_{CY}) \geq H(V_X)$, $H(V_{CY}) \geq H(V_Y)$ and $H(V_{CY}) \geq H(V_{CX})$.
 
\begin{eqnarray}
M_{CY} = 2^{K h_{Y}}
\label{theorem4_proof_6}
\end{eqnarray}

In the same way as theorem 2 and 3, the codewords $W_X$ and $W_Y$ and their keys $W_{kX}$ and $W_{kY}$ are now defined:

\begin{eqnarray}
W_X = (W_{X1} \oplus W_{kCY}, W_{X2}  \oplus W_{kCY}, W_{CX}  \oplus W_{kCY})
\label{theorem4_proof_7}
\end{eqnarray}

\begin{eqnarray}
W_Y = (W_{Y1} \oplus W_{kCY}, W_{Y2}  \oplus W_{kCY}, W_{CY})
\label{theorem4_proof_8}
\end{eqnarray}

\begin{eqnarray}
W_{kY} = (W_{kCY})
\label{theorem4_proof_10}
\end{eqnarray}

where $W_\alpha \in I_{M_\alpha} = \{0, 1, \ldots, M_\alpha - 1\}$. The wiretapper will not know $W_{X1}$, $W_{X2}$ and $W_{CX}$ from $W_X$ and $W_{Y1}$, $W_{Y2}$ and $W_{CY}$ from $W_Y$ as these are protected by the key ($W_{kCY}$.

\begin{eqnarray}
\frac{1}{K} \log M_X + \frac{1}{K} \log M_Y  & = & \frac{1}{K} (\log M_{X1} + \log M_{X2}  \nonumber
\\ & + &  \log M_{CX}) + \frac{1}{K} (\log M_{Y1}  \nonumber
\\ & + &  \log M_{Y2} +  \log M_{CY}) \nonumber 
\\ & \le & H(X|Y) + H(Y|X) 				\nonumber
\\ & + & I(X;Y) + \epsilon_0 \nonumber
\\ & = & H(X,Y)  		\nonumber
\\ & \le & R_X + R_Y
\label{theorem4_proof_11}
\end{eqnarray}

\begin{eqnarray}
\frac{1}{K} \log M_{kX} & = & \frac{1}{K} \log M_{CY}	\nonumber
\\ & = & h_{Y}			\label{num4.2}
\\ & \geq & h_X - \epsilon_0  \label{num4.3}
\\ &  & R_{kX}
\label{theorem4_proof_13}
\end{eqnarray}

where \eqref{num4.2} comes from \eqref{theorem4_proof_6} and \eqref{num4.3} comes form the consideration stated at the beginning of this proof.

\begin{eqnarray}
\frac{1}{K} \log M_{kY} & = & \frac{1}{K} \log M_{CY}	\nonumber
\\ & = & h_{XY}			\label{num5.1}
\\ & \le & R_{kY}
\label{theorem4_proof_14}
\end{eqnarray}
where \eqref{num5.1} comes from \eqref{theorem4_proof_6}.

The security levels thus result:
\begin{eqnarray}
\frac{1}{K} H(X^K|W_X, W_Y) & = & \frac{1}{K} H(X|W_{X1} \oplus W_{kCY}, 
\\ && W_{X2} \oplus W_{kCY}, W_{CX} \oplus W_{kCY},			\nonumber
\\ && W_{Y1} \oplus W_{kCY}, W_{Y2} \oplus W_{kCY},	\nonumber
\\ &&  W_{CY}) 				\nonumber
\\ & = & H(X^K)	\label{num6.1}
\\ & \ge & h_X - \epsilon^{'}_0
\label{theorem4_proof_16}
\end{eqnarray}

where \eqref{num6.1} holds because $W_{X1}$, $W_{X2}$, $W_{CX}$, $W_{Y1}$, $W_{Y2}$ are covered by key $W_{CY}$ and $W_{CY}$ is covered by an existing random number key. Equations \eqref{lemma1_1} - \eqref{lemma1_7} imply that $W_{X1}$, $W_{X2}$, $W_{Y1}$ and $W_{Y2}$ have almost no redundancy and they are mutually independent.

Similarly, 
\begin{eqnarray}
\frac{1}{K} H(Y^K|W_X, W_Y) \geq h_Y - \epsilon^{'}_0
\label{theorem4_proof_17}
\end{eqnarray}

Therefore ($R_X$, $R_Y$, $R_{kX}$, $R_{kY}$, $h_{XY}$, $h_{XY}$) is admissible from \eqref{theorem2_proof_11} -  \eqref{theorem2_proof_17}.

Next the case where: $H(V_{CY}) < H(V_X)$, $H(V_{CY}) < H(V_Y)$ and $H(V_{CY}) < H(V_{CX})$ is considered. Here, there are shorter length keys used in addition to the key provided by $W_{CY}$ in order to make the key lengths required by the individual portions. For example the key $W_{k1}$ comprises $W_{kCY}$ and a short key $W_1$, which together provide the length of $W_{X1}$.
The codewords $W_X$ and $W_Y$ and their keys $W_{kX}$ and $W_{kY}$ are now defined:

\begin{eqnarray}
W_X = (W_{X1} \oplus W_{k1}, W_{X2}  \oplus W_{k2}, W_{CX}  \oplus W_{k3})
\label{theorem4_proof_7.1.1}
\end{eqnarray}

\begin{eqnarray}
W_Y = (W_{Y1} \oplus W_{k4}, W_{Y2}  \oplus W_{k5}, W_{CY})
\label{theorem4_proof_8.1.1}
\end{eqnarray}

\begin{eqnarray}
W_{kX} = (W_{k1}, W_{k2}, W_{k3})
\label{theorem4_proof_10.1.1}
\end{eqnarray}

\begin{eqnarray}
W_{kY} = (W_{k4}, W_{k5})
\label{theorem4_proof_10.1.2}
\end{eqnarray}

where $W_\alpha \in I_{M_\alpha} = \{0, 1, \ldots, M_\alpha - 1\}$. The wiretapper will not know $W_{X1}$, $W_{X2}$ and $W_{CX}$ from $W_X$ and $W_{Y1}$, $W_{Y2}$ and $W_{CY}$ from $W_Y$ as these are protected by the key ($W_{kCY}$.

In this case, $R_X$, $R_Y$, $R_{kX}$ and $R_{kY}$ satisfy that

\begin{eqnarray}
\frac{1}{K} \log M_X + \frac{1}{K} \log M_Y  & = & \frac{1}{K} (\log M_{X1} \nonumber
\\ & + & \log M_{X2} +  \log M_{CX}) \nonumber
\\ & + & \frac{1}{K} (\log M_{Y1} + \log M_{Y2} +  \log M_{CY}) \nonumber
\\ & \le & H(X|Y) + H(Y|X) + I(X;Y) + \epsilon_0				\nonumber
\\ & = & H(X,Y)  		\nonumber
\\ & \le & R_X + R_Y
\label{theorem4_proof_11.1.1}
\end{eqnarray}

\begin{eqnarray}
\frac{1}{K} \log M_{kX} & = & \frac{1}{K} [\log M_{k1} + \log M_{k2} + \log M_{k3}]	\nonumber
\\ & = & \log M_{kCY} + \log M_{1} +	\log M_{kCY} \nonumber
\\ & + & \log M_{2} + \log M_{kCY} + \log M_{3} \nonumber
\\ & = & 3 \log M_{kCY} + \log M_{1} + \log M_{2} + \log M_{3} \nonumber
\\ & \geq & 3 h_{Y} - \epsilon_0  \label{num335.1}
\\ & \geq & 3 h_{X} - \epsilon_0 \nonumber
\\ & \geq & h_{X}
\label{theorem4_proof_13.1.1}
\end{eqnarray}

where \eqref{num335.1} results from \eqref{theorem4_proof_6} and the result is from the consideration at the beginning of this proof.

\begin{eqnarray}
\frac{1}{K} \log M_{kY} & = & \frac{1}{K} [\log M_{k3} + \log M_{k4} + \log M_{kCY}]	\nonumber
\\ & = & \log M_{kCY} + \log M_{3} +	\log M_{kCY} \nonumber
\\ & + &  \log M_{4} + \log M_{kCY}  \nonumber
\\ & = & 3 \log M_{kCY} + \log M_{3} + \log M_{4} \nonumber
\\ & \geq & 3 h_{Y} - \epsilon_0 \label{num336.5}
\\ & \geq & h_{Y}
\label{theorem4_proof_14.1.1}
\end{eqnarray}

where \eqref{num336.5} results from \eqref{theorem4_proof_6}.

The security levels thus result:
\begin{eqnarray}
\frac{1}{K} H(X^K|W_X, W_Y) & = & \frac{1}{K} H(X|W_{X1} \oplus W_{k1}, 
\\ && W_{X2} \oplus W_{k2}, W_{CX} \oplus W_{k3},			\nonumber
\\ && W_{Y1} \oplus W_{k4}, W_{Y2} \oplus W_{k5},	\nonumber
\\ &&  W_{CY}) 				\nonumber
\\ & = & H(X^K)	\label{num6.1.1.1}
\\ & \ge & h_X - \epsilon^{'}_0
\label{theorem4_proof_16.1.1}
\end{eqnarray}

where \eqref{num6.1.1.1} holds because $W_{X1}$, $W_{X2}$, $W_{CX}$, $W_{Y1}$, $W_{Y2}$ are covered by key $W_{CY}$ and some shorter length key and $W_{CY}$ is covered by an existing random number key. 

Similarly, 
\begin{eqnarray}
\frac{1}{K} H(Y^K|W_X, W_Y) \geq h_Y - \epsilon^{'}_0
\label{theorem4_proof_17.1.1}
\end{eqnarray}

Therefore ($R_X$, $R_Y$, $R_{kX}$, $R_{kY}$, $h_{XY}$, $h_{XY}$) is admissible from \eqref{theorem4_proof_11.1.1} -  \eqref{theorem4_proof_17.1.1}.

The region indicated for $\mathcal{R_4}$ is derived from this region for $\mathcal{R_3}$, when $h_X = 0$.

\end{proof}

\begin{proof}[Proof of Theorem 5]
As before, $V_{CY}$ may be used as a key, however here we use $V_{CX}$ as the key in this proof to show some variation. 

Now the consideration for the security levels is that $h_X \geq h_Y$ because $X$ contains the key that is used for masking.
Suppose that ($R_X$, $R_Y$, $R_{KX}$, $R_{KY}$) $\in$ 
$\mathcal{R}_5$. From \eqref{theorem3} 
\begin{eqnarray}
&& R_X \geq H(X^K|Y^K)  				\nonumber
\\&& R_Y  \geq H(Y^K|X^K) 				\nonumber
\\&& R_X + R_Y  \geq H(X^K, Y^K)
\label{theorem5_proof_1}
\end{eqnarray}

\begin{eqnarray}
R_{kX} \geq h_{X}, R_{kY} \geq h_{Y} 
\label{theorem5_proof_2}
\end{eqnarray}

Assuming a key exists for $V_{CX}$. For the first case, consider the following: $H(V_{CX}) \geq H(V_X)$, $H(V_{CX}) \geq H(V_Y)$ and $H(V_{CX}) \geq H(V_{CX})$.
 
\begin{eqnarray}
M_{CX} = 2^{K h_{X}}
\label{theorem5_proof_6}
\end{eqnarray}

The codewords $W_X$ and $W_Y$ and their keys $W_{kX}$ and $W_{kY}$ are now defined:

\begin{eqnarray}
W_X = (W_{X1} \oplus W_{kCX}, W_{X2}  \oplus W_{kCX}, W_{CX})
\label{theorem5_proof_7}
\end{eqnarray}

\begin{eqnarray}
W_Y = (W_{Y1} \oplus W_{kCX}, W_{Y2}  \oplus W_{kCX}, W_{CY}  \oplus W_{kCX})
\label{theorem5_proof_8}
\end{eqnarray}

\begin{eqnarray}
W_{kX} = (W_{kCX})
\label{theorem5_proof_10}
\end{eqnarray}

where $W_\alpha \in I_{M_\alpha} = \{0, 1, \ldots, M_\alpha - 1\}$. The wiretapper will not know $W_{X1}$, $W_{X2}$ and $W_{CX}$ from $W_X$ and $W_{Y1}$, $W_{Y2}$ and $W_{CY}$ from $W_Y$ as these are protected by the key $W_{kCX}$ and $W_{kCX}$ is protected by a random number key.

\begin{eqnarray}
\frac{1}{K} \log M_X + \frac{1}{K} \log M_Y  & = & \frac{1}{K} (\log M_{X1} + \log M_{X2}  \nonumber
\\ & + &  \log M_{CX}) + \frac{1}{K} (\log M_{Y1}  \nonumber
\\ & + &  \log M_{Y2} +  \log M_{CY}) \nonumber 
\\ & \le & H(X|Y) + H(Y|X) 				\nonumber
\\ & + & I(X;Y) + \epsilon_0 \nonumber
\\ & = & H(X,Y)  		\nonumber
\\ & \le & R_X + R_Y
\label{theorem5_proof_11}
\end{eqnarray}

\begin{eqnarray}
\frac{1}{K} \log M_{kX} & = & \frac{1}{K} \log M_{CX}	\nonumber
\\ & \geq & h_X - \epsilon_0  \label{num6.31}
\\ &  & R_{kX}
\label{theorem5_proof_13}
\end{eqnarray}

where \eqref{num6.31} comes from \eqref{theorem5_proof_6}.

\begin{eqnarray}
\frac{1}{K} \log M_{kY} & = & \frac{1}{K} \log M_{CX}	\nonumber
\\ & = & h_{X}			\label{num6.32}
\\ & \geq & h_Y \label{num6.32}
\\ & \le & R_{kY}
\label{theorem5_proof_14}
\end{eqnarray}
where \eqref{num6.32} comes from \eqref{theorem5_proof_6} and \eqref{num6.32} comes form the consideration stated at the beginning of this proof.

The security levels thus result:
\begin{eqnarray}
\frac{1}{K} H(X^K|W_X, W_Y) & = & \frac{1}{K} H(X|W_{X1} \oplus W_{kCX}, \nonumber
\\ && W_{X2} \oplus W_{kCX}, W_{CX}, W_{CY} \oplus W_{kCX},			\nonumber
\\ && W_{Y1} \oplus W_{kCX}, W_{Y2} \oplus W_{kCX}) 				\nonumber
\\ & = & H(X^K)	\label{num16.1}
\\ & \ge & h_X - \epsilon^{'}_0
\label{theorem5_proof_16}
\end{eqnarray}

where \eqref{num6.1} holds because $W_{X1}$, $W_{X2}$, $W_{CX}$, $W_{Y1}$, $W_{Y2}$ are covered by key $W_{CY}$ and $W_{CY}$ is covered by an existing random number key. Equations \eqref{lemma1_1} - \eqref{lemma1_7} imply that $W_{X1}$, $W_{X2}$, $W_{Y1}$ and $W_{Y2}$ have almost no redundancy and they are mutually independent.

Similarly, 
\begin{eqnarray}
\frac{1}{K} H(Y^K|W_X, W_Y) \geq h_Y - \epsilon^{'}_0
\label{theorem5_proof_17}
\end{eqnarray}

Therefore ($R_X$, $R_Y$, $R_{kX}$, $R_{kY}$, $h_{XY}$, $h_{XY}$) is admissible from \eqref{theorem2_proof_11} -  \eqref{theorem2_proof_17}.

Next the case where: $H(V_{CX}) < H(V_X)$, $H(V_{CX}) < H(V_Y)$ and $H(V_{CX}) < H(V_{CX})$ is considered. Here, there are shorter length keys used in addition to the key provided by $W_{CX}$ in order to make the key lengths required by the individual portions. For example the key $W_{k1}$ comprises $W_{kCX}$ and a short key $W_1$, which together are the length of $W_{X1}$.
The codewords $W_X$ and $W_Y$ and their keys $W_{kX}$ and $W_{kY}$ are now defined:

\begin{eqnarray}
W_X = (W_{X1} \oplus W_{k1}, W_{X2}  \oplus W_{k2}, W_{CX})
\label{theorem5_proof_7.1.1}
\end{eqnarray}

\begin{eqnarray}
W_Y = (W_{CY} \oplus W_{k3}, W_{Y1} \oplus W_{k4}, W_{Y2}  \oplus W_{k5})
\label{theorem5_proof_8.1.1}
\end{eqnarray}

\begin{eqnarray}
W_{kX} = (W_{k1}, W_{k2}, W_{k3})
\label{theorem5_proof_10.1.1}
\end{eqnarray}

\begin{eqnarray}
W_{kY} = (W_{k4}, W_{k5})
\label{theorem5_proof_10.1.2}
\end{eqnarray}

where $W_\alpha \in I_{M_\alpha} = \{0, 1, \ldots, M_\alpha - 1\}$. The wiretapper will not know $W_{X1}$, $W_{X2}$ and $W_{CX}$ from $W_X$ and $W_{Y1}$, $W_{Y2}$ and $W_{CY}$ from $W_Y$ as these are protected by the key ($W_{kCY}$.

In this case, $R_X$, $R_Y$, $R_{kX}$ and $R_{kY}$ satisfy that

\begin{eqnarray}
\frac{1}{K} \log M_X + \frac{1}{K} \log M_Y  & = & \frac{1}{K} (\log M_{X1} \nonumber
\\ & + & \log M_{X2} +  \log M_{CX}) \nonumber
\\ & + & \frac{1}{K} (\log M_{Y1} + \log M_{Y2} \nonumber
\\ & + &  \log M_{CY}) \nonumber
\\ & \le & H(X|Y) + H(Y|X) + I(X;Y)  \nonumber
\\ & + & \epsilon_0				\nonumber
\\ & = & H(X,Y)  		\nonumber
\\ & \le & R_X + R_Y
\label{theorem5_proof_11.1.1}
\end{eqnarray}

\begin{eqnarray}
\frac{1}{K} \log M_{kX} & = & \frac{1}{K} [\log M_{k1} + \log M_{k2} + \log M_{k3}]	\nonumber
\\ & = & \log M_{kCX} + \log M_{1} +	\log M_{kCX}  \nonumber
\\ & + & \log M_{2} + \log M_{kCX} + \log M_{3} \nonumber
\\ & = & 3 \log M_{kCX} + \log M_{1} + \log M_{2} + \log M_{3} \nonumber
\\ & \geq & 3 h_{X} - \epsilon_0  \label{num3335.1}
\\ & \geq & h_{X}
\label{theorem5_proof_13.1.1.1}
\end{eqnarray}

where \eqref{num3335.1} results from \eqref{theorem5_proof_6}.
\begin{eqnarray}
\frac{1}{K} \log M_{kY} & = & \frac{1}{K} [\log M_{k3} + \log M_{k4} + \log M_{kCY}]	\nonumber
\\ & = & \log M_{kCX} + \log M_{3} +	\log M_{kCX} \nonumber
\\ & + & \log M_{4} + \log M_{kCX} \nonumber
\\ & = & 3 \log M_{kCX} + \log M_{3} + \log M_{4} \nonumber
\\ & \geq & 3 h_{X} - \epsilon_0  \label{num33335.1}
\\ & \geq & 3 h_{Y} - \epsilon_0 \label{num336.5}
\\ & \geq & h_{Y}
\label{theorem5_proof_14.1.1}
\end{eqnarray}

where \eqref{num33335.1} results from \eqref{theorem5_proof_6} and \eqref{num336.5} results from the consideration at the beginning of this proof.

The security levels thus result:
\begin{eqnarray}
\frac{1}{K} H(X^K|W_X, W_Y) & = & \frac{1}{K} H(X|W_{X1} \oplus W_{k1}, W_{X2} \oplus W_{k2}, \nonumber
\\ && W_{CX}, W_{CY} \oplus W_{k3},			\nonumber
\\ && W_{Y1} \oplus W_{k4}, W_{Y2} \oplus W_{k5},	\nonumber
\\ &&  W_{CY}) 				\nonumber
\\ & = & H(X^K)	\label{num6.1.1.1.1}
\\ & \ge & h_X - \epsilon^{'}_0
\label{theorem5_proof_16.1.1.1}
\end{eqnarray}

where \eqref{num6.1.1.1.1} holds because $W_{X1}$, $W_{X2}$, $W_{CY}$, $W_{Y1}$, $W_{Y2}$ are covered by key $W_{CX}$ and some shorter length key and $W_{CX}$ is covered by an existing random number key. 

Similarly, 
\begin{eqnarray}
\frac{1}{K} H(Y^K|W_X, W_Y) \geq h_Y - \epsilon^{'}_0
\label{theorem5_proof_17.1.1}
\end{eqnarray}

Therefore ($R_X$, $R_Y$, $R_{kX}$, $R_{kY}$, $h_{XY}$, $h_{XY}$) is admissible from \eqref{theorem5_proof_11.1.1} -  \eqref{theorem5_proof_17.1.1}.

\end{proof}

\subsection{Converse parts}
From Slepian-Wolf's theorem we know that the channel rate must satisfy $R_X \geq H(X|Y)$, $R_Y \geq H(Y|X)$ and $R_X + R_Y \geq H(X,Y)$ to achieve a low error probability when decoding.
Hence, the key rates are considered in this subsection. 
\\
\textit{Converse part of Theorem 2:}
\begin{eqnarray}
R_{kX} & \geq & \frac{1}{K} log M_{kX} - \epsilon		\nonumber
\\ & \geq & \frac{1}{K} H(W_{kX}) - \epsilon			\nonumber
\\ & \geq & \frac{1}{K} H(W_{kX|W_1}) - \epsilon			\nonumber
\\ & = & \frac{1}{K} [H(W_{kX}) - I(W_{kX}; W_1)] - \epsilon		\nonumber
\\ & = & \frac{1}{K} [H((W_{kX|X, Y, W_1}) + I(W_{kX}; W_1) 		\nonumber
\\ & + & I(W_{kX};X|Y, W_1) + I(X, Y, W_{kX}|W_1) 		\nonumber
\\ & + & I(Y, W_{kX}|X, W_1) - I(W_{kX}; W_1)] - \epsilon		\nonumber
\\ & = & \frac{1}{K} [H(X, Y|W_1) - H(X,Y|W_1, W_{kX})] - \epsilon		\nonumber
\\ & \geq & h_{XY} - \frac{1}{K} H(X,Y|W_1, W_{kX})  - \epsilon  \label{conv_1}
\\ & = & h_{XY} - H(V_{CY})  - \epsilon  		\nonumber
\\ & = & h_{XY}  - \epsilon
\end{eqnarray}

where \eqref{conv_1} results from equation \eqref{cond8.1}. Here, we consider the extremes of $H(V_{CY})$ in order to determine the limit for $R_{kX}$. When this quantity is minimum then we are able to achieve the maximum bound of $h_{XY}$.

\begin{eqnarray}
R_{kY} & \geq & \frac{1}{K} log M_{kY} - \epsilon		\nonumber
\\ & \geq & \frac{1}{K} H(W_{kY}) - \epsilon			\nonumber
\\ & \geq & \frac{1}{K} H(W_{kY|W_2}) - \epsilon		\nonumber
\\ & = & \frac{1}{K} [H(W_{kY}) - I(W_{kY}; W_2)] - \epsilon		\nonumber
\\ & = & \frac{1}{K} [H((W_{kY|X, Y, W_2}) + I(W_{kY}; W_2) 		\nonumber
\\ & + & I(W_{kY};X|Y, W_2) + I(X, Y, W_{kY}|W_2) 				\nonumber
\\ & + & I(Y, W_{kY}|X, W_2) - I(W_{kY}; W_2)] - \epsilon			\nonumber
\\ & = & \frac{1}{K} [H(X, Y|W_2) - H(X,Y|W_2, W_{kY})] - \epsilon		\nonumber
\\ & \geq & h_{XY} - \frac{1}{K} H(X,Y|W_2, W_{kY})  - \epsilon  \label{conv_2}
\\ & = & h_{XY} - H(V_{CX})  - \epsilon   	\nonumber
\\ & = & h_{XY}  - \epsilon
\end{eqnarray}

where \eqref{conv_2} results from equation \eqref{cond9}. Here, we consider the extremes of $H(V_{CX})$ in order to determine the limit for $R_{kY}$. When this quantity is minimum then we are able to achieve the maximum bound of $h_{XY}$.

\textit{Converse part of Theorem 3:}
\begin{eqnarray}
R_{kX} & \geq & \frac{1}{K} log M_{kX} - \epsilon		\nonumber
\\ & \geq & \frac{1}{K} H(W_{kX}) - \epsilon		\nonumber
\\ & \geq & \frac{1}{K} H(W_{kX|W_1}) - \epsilon		\nonumber
\\ & = & \frac{1}{K} [H(W_{kX}) - I(W_{kX}; W_1)] - \epsilon		\nonumber
\\ & = & \frac{1}{K} [H((W_{kX|X, W_1}) + I(W_{kX}; W_1) 		\nonumber
\\ & + & I(X, W_{kX}|W_1) - I(W_{kX}; W_1)] - \epsilon		\nonumber
\\ & \geq & \frac{1}{K} I(X, W_{kX}|W_1) - \epsilon		\nonumber
\\ & = & \frac{1}{K} [H(X|W_1) - H(X|W_1, W_{kX})] - \epsilon	\nonumber
\\ & \geq & h_{X} - H(V_{CY})  - \epsilon  \label{conv_3}
\\ & = & h_{X}  - \epsilon
\end{eqnarray}

where \eqref{conv_3} results from \eqref{cond7}. Here, we consider the extremes of $H(V_{CY})$ in order to determine the limit for $R_{kX}$. When this quantity is minimum then we are able to achieve the maximum bound of $h_{X}$.

\begin{eqnarray}
R_{kY} & \geq & \frac{1}{K} log M_{kY} - \epsilon	\nonumber
\\ & \geq & \frac{1}{K} H(W_{kY}) - \epsilon		\nonumber
\\ & \geq & \frac{1}{K} H(W_{kY|W_2}) - \epsilon		\nonumber
\\ & = & \frac{1}{K} [H(W_{kY}) - I(W_{kY}; W_2)] - \epsilon		\nonumber
\\ & = & \frac{1}{K} [H((W_{kY|Y, W_2}) + I(W_{kY}; W_2) 		\nonumber
\\ & + & I(X, W_{kY}|W_2) - I(W_{kY}; W_2)] - \epsilon		\nonumber
\\ & \geq & \frac{1}{K} I(Y, W_{kY}|W_2) - \epsilon				\nonumber
\\ & = & \frac{1}{K} [H(Y|W_2) - H(Y|W_2, W_{kY})] - \epsilon		\nonumber
\\ & \geq & h_{Y} - H(V_{CX})  - \epsilon  \label{conv_4}
\\ & = & h_{Y}  - \epsilon
\end{eqnarray}

where \eqref{conv_3} results from \eqref{cond8}. Here, we consider the extremes of $H(V_{CX})$ in order to determine the limit for $R_{kY}$. When this quantity is minimum then we are able to achieve the maximum bound of $h_{Y}$.

Since theorems 4-5 also have key rates of $h_X$ and $h_Y$ for $X$ and $Y$ respectively we can use the same methods to prove the converse.

\section{Scheme for multiple sources}

The two correlated source model presented in Section II is generalised even further, and now concentrates on multiple correlated sources transmitting syndromes across multiple wiretapped links. This new approach represents a network scenario where there are many sources and one receiver. We consider the information leakage for this model for Slepian-Wolf coding and thereafter consider the Shannon's cipher system representation.

\subsection{Information leakage using Slepian-Wolf coding}

\begin{figure}[ht]
\centering
\includegraphics [scale = 0.7]{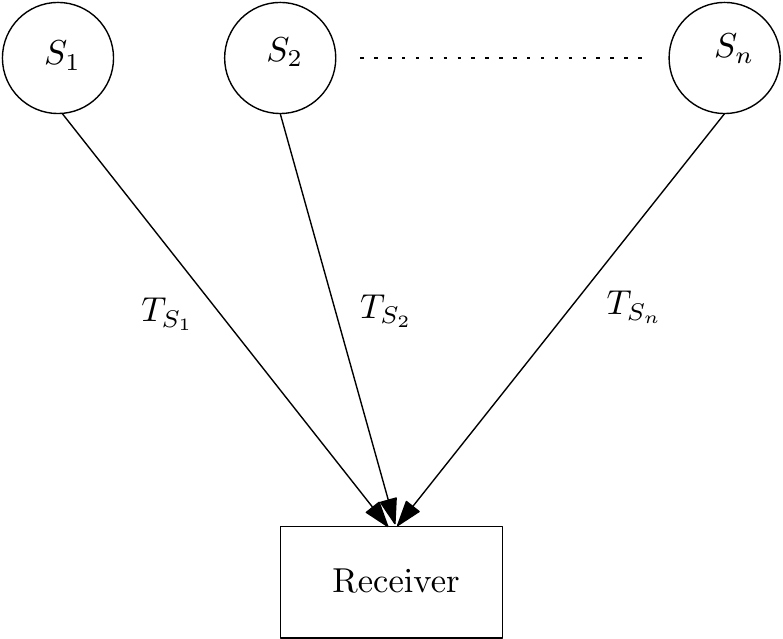}
\caption{Extended generalised model}
\label{fig:extended}
\end{figure}

Here, Figure~\ref{fig:extended} gives a pictorial view of the new extended model for multiple correlated sources.

Consider a situation where there are many sources, which are part of the ${\bf S}$ set:
 
\begin{eqnarray}
{\bf S} = \{S_{1}, S_{2}, \ldots, S_{n}\} \nonumber
\end{eqnarray}
where $i$ represents the $i$th source ($i = 1,\ldots, n$) and there are $n$ sources in total. Each source may have some correlation between some other source and all sources are part of a binary alphabet. There is one receiver that is responsible for performing decoding. The syndrome for a source $S_i$ is represented by $T_{S_i}$, which is part of the same alphabet as the sources. 

The entropy of a source is given by a combination of a specific conditional entropy and mutual information. In order to present the entropy we first define the following sets:

\begin{itemize}
\item [-] The set, ${\bf S}$ that contains all sources: ${\bf S} = \{S_1, S_2,\ldots, S_n\}$. 
\item [-] The set, ${\bf S}_t$ that contains $t$ unique elements from ${\bf S}$ and ${\bf S}_t$ $\subseteq$ $\bf{S}$, ${S}_i \in {\bf S}_t$, ${\bf S}_t \cup {\bf S}_t^c$ $=$ $\bf{S}$  and $|{\bf S}_t|$ $= t$ 
\end{itemize}

Here, $H(S_i)$ is obtained as follows:

\begin{eqnarray}
H(S_i) = H(S_i|{\bf S}_{\backslash S_i}) + \displaystyle\sum_{t=2}^{n} (-1)^{t-1} \displaystyle\sum_{\text{all possible ${\bf S}_t$}}^{} I({\bf S}_t|{\bf S}_t^c)
\label{entropy}
\end{eqnarray}

Here, $n$ is the number of sources, $ H(S_i|{\bf S}_{\backslash S_i})$ denotes the conditional entropy of the source $S_i$ given $S_i$ subtracted from the set ${\bf S}$ and $I({\bf S}_t|{\bf S}_t^c)$ denotes the mutual information between all sources in the subset ${\bf S}_t$ given the complement of ${\bf S}_t$.
In the same way as for two sources, the generalised probabilities and entropies can be developed. It is then possible to decode the source message for source $S_i$ by receiving all components related to $S_i$. This gives rise to the following inequality for $H(S_i)$ in terms of the sources:
\begin{eqnarray}
H(S_i|{ {\bf S}_{\backslash S_i}}) & + & \displaystyle\sum_{t=2}^{n} (-1)^{t-1}	\nonumber \displaystyle\sum_{\text{all possible ${\bf S}_t$}}^{} I({\bf S}_t|{\bf S}_t^c) 
\\ & \le & H(S_i) + \delta
\label{3source_2}
\end{eqnarray}

In this type of model information from multiple links need to be gathered in order to determine the transmitted information for one source. Here, the common information between sources is represented by the $I({\bf S}_t|{\bf S}_t^c)$ term. The portions of common information sent by each source can be determined upfront and is an arbitrary allocation in our case. For example in a three source model where $X$, $Y$ and $Z$ are the correlated sources, the common information shared with $X$ and the other sources is represented as: $I(X;Y|Z)$ and $I(X;Z|Y)$. Each common information portion is divided such that the sources having access to it are able to produce a portion of it themselves. The common information $I(X;Y|Z)$ is divided into $V_{CX1}$ and $V_{CY1}$ where the former is the common information between $X$ and $Y$, produced by $X$ and the latter is the common information between $X$ and $Y$, produced by $Y$. Similarly, $I(X;Z|Y)$ consists of two common information portions, $V_{CX2}$ and $V_{CZ1}$ produced by $X$ and $Z$ respectively. 

As with the previous model for two correlated sources, since wiretapping is possible there is a need to develop the information leakage for the model. The information leakages for this multiple source model is indicated in \eqref{remark1} and \eqref{remark2}. 

\textit{Remark 1:} The leaked information for a source $S_i$ given the transmitted codewords $T_{S_i}$, is given by:
\begin{eqnarray}
L^{S_i}_{T_{S_i}} = I(S_i ; T_{S_i})
\label{remark1}
\end{eqnarray}

Since we use the notion that the information leakage is the conditional entropy of the source given the transmitted information subtracted from the source's uncertainty (i.e $H(S_i) - H(S_i| T_{S_i})$), the proof for \eqref{remark1} is trivial. Here, we note that the common information is the minimum amount of information leaked. Each source is responsible for transmitting its own private information and there is a possibility that this private information may also be leaked. The maximum leakage for this case is thus the uncertainty of the source itself, $H(S_i)$.

We also consider the information leakage for a source $S_i$ when another source $S_{j_(j \neq i)}$ has transmitted information. This gives rise to Remark 2.
\\
\textit{Remark 2:} The leaked information for a source $S_i$ given the transmitted codewords $T_{S_j}$, where $i \neq j$ is:
\begin{eqnarray}
L^{S_i}_{T_{S_j}} & = & H(S_i) - H(S_i|T_{S_j})	\nonumber
\\ & = & H(S_i) - [H(S_i) - I(S_i; T_{S_j})]	\nonumber
\\ & = & I(S_i ; T_{S_j})
\label{remark2}
\end{eqnarray}

The information leakage for a source is determined based on the information transmitted from any other channel using the common information between them. The private information is not considered as it is transmitted by each source itself and can therefore not be obtained from an alternate channel. Remark 2 therefore gives an indication of the maximum amount of information leaked for source $S_i$, with knowledge of the syndrome $T_{S_j}$. 

These remarks show that the common information can be used to quantify the leaked information. The common information provides information for more than one source and is therefore susceptible to leaking information about more than one source should it be compromised. This subsection gives an indication of the information leakage for the new generalised multiple correlated sources model when a source's syndrome and other syndromes are wiretapped.

\subsection{Information leakage for Shannon's cipher system}

This subsection details a novel masking method to minimize the key length and thereafter builds this multiple correlated source model on Shannon's cipher system. 

The new masking method encompasses masking the conditional entropy portion with a mutual information portion. By masking, certain information is hidden and it becomes more difficult to obtain the information that has been masked. Masking can typically be done using random numbers, however we eliminate the need for random numbers that represent keys and rather use a common information to mask with. 

We make the following assumptions:
\begin{itemize}
\item
The capacity of each link cannot be exhausted using this method.
\item
A common information is used to mask certain private information and can be used to mask multiple times. Further, private information that needs to be masked always exists in this method.
\end{itemize}
The allocation of common information for transmission are done on an arbitrary basis. The objective of this subsection is to minimize the key lengths while achieving perfect secrecy. 

The private information for source $i$ is given by $H(S_i|{ {\bf S}_{\backslash S_i}})$ according to \eqref{entropy}, which is called $W_{S_i}$ and the common information associated with source $S_i$ is given by $W_{CS_i}$. First, choose a common information with which to mask. Then we take a part of  $W_{S_i}$, i.e.  ${W_{S_i}}^{'}$, that has entropy equal to $H(W_{CS_i})$, and mask as follows: 
\begin{eqnarray}
W_{S_i}^{'} \oplus W_{CS_i}
\label{masking}
\end{eqnarray}

When the two sequences are xor'ed the result is a single sequence that may look different to the originals. We then transmit the masked portion instead of the $W_{S_i}^{'}$ portion when transmitting $W_{S_i}$, thus providing added security. 
This brings in the interesting factor that there are many possibilities for a specific mutual information to mask conditional entropy portions. For example when considering  three sources as before, it is possible to mask the private information for $X$, $Y$ and $Z$ with the common portion $I(X;Y;Z)$. If $Y$ is secure then this common information can be transmitted along $Y$'s channel, ensuring the information is kept secure. The ability to mask using the common information is a unique and interesting feature of this new model for multiple correlated sources. The underlying principle is that the secure link should transmit more common information after transmitting their private information. 

We find that the lower bound for the channel rate when the masking approach is used is given by:
\begin{eqnarray}
R_i^M \geq H(S_1, \ldots, S_n) - \displaystyle\sum_{t=2}^{n} \displaystyle\sum_{\text{all possible ${\bf S}_t$}} (t -1)  I({\bf S}_t|{\bf S}_t^c)
\end{eqnarray}
where $R_i^M$ is the $i$th channel rate when masking is used. 

The method works theoretically but may result in some concern practically as there may be a security compromise when common information is sent across non secure links. We see that if the $ W_{CS_i}$ component used for masking has been compromised then the private portion it has masked will also be compromised. A method to overcome this involves using two common information parts for masking. Equation \eqref{masking} representing the masking would become:
\begin{eqnarray}
W_{S_i}^{'} \oplus W_{CS_i} \oplus W_{CS_j}
\label{masking2}
\end{eqnarray}

where $i \neq j$ and both $W_{CS_i}$ and $W_{CS_j}$ are common information associated with source $S_i$. This way, if only $W_{CS_j}$ is compromised then $W_{S_i}$ is not compromised as it is still protected by $W_{CS_i}$. Here, combinations of common information are used to increase the security. The advantage with \eqref{masking2} is that keys may be reused because common information may be shared by more than one source. Further, the method will not result in an increase in key length.

The Shannon's cipher system for this multiple source model is now presented in order to determine the rate regions for perfect secrecy. The multiple sources each have their own encoder and there is a universal decoder. Each source has an encoder represented by:
\begin{eqnarray}
E_i : \mathcal{S} \times I_{W_{S_i}} & \rightarrow & I_{W_{CS_i}} =  \{0, 1, \ldots, W_{S_i} - 1\} \nonumber 
\\ && I_{W_{CS_i}} =  \{0, 1, \ldots, W_{CS_i} - 1\}
\label{ms_xencoder_fcn}
\end{eqnarray}
where $I_{MPi}$ is the alphabet representing the private portion for source $S_i$ and $I_{MCi}$ is the alphabet representing the common information for source $S_i$.
The decoder at the receiver is defined as:
\begin{eqnarray}
D : (I_{W_{S_i}}, I_{W_{CS_i}})  & \times &  I_{Mk} \rightarrow \mathcal{S}
\end{eqnarray}

The encoder and decoder mappings are below:
\begin{eqnarray}
W_i = F_{E_i} (S_i, W_{ki})
\end{eqnarray}

\begin{eqnarray}
\widehat{S_i} = F_{D_i} (W_i, W_{ki}, W_{\{{kp}\}})
\end{eqnarray}

where $p = 1, \ldots, n$, $p \neq i$ and $W_{\{{kp}\}}$ represents the set of common information required to find $S_i$, and $\widehat{S_i}$ is the decoded output. 

The following conditions should be satisfied for the general cases:

\begin{eqnarray}
\frac{1}{K}\log W_{S_i} \le R_i +\epsilon
\label{ms_cond1}
\end{eqnarray}

\begin{eqnarray}
\frac{1}{K}\log M_{ki} \le R_{ki} +\epsilon
\label{ms_cond2}
\end{eqnarray}

\begin{eqnarray}
\text {Pr} \{\widehat{S_i} \neq S_i\} \le \epsilon
\label{ms_cond3}
\end{eqnarray}

\begin{eqnarray}
\frac{1}{K} H(S_i|W_i) \le h_i - \epsilon
\label{ms_cond7}
\end{eqnarray}

\begin{eqnarray}
\frac{1}{K} H(S_j|W_i) \le h_j - \epsilon
\label{ms_cond8}
\end{eqnarray}

where $R_i$ is the the rate of source $S_i$'s channel and $R_{k_{i}}$ is the key rate of $S_i$. The security levels, for source $i$ and any other source $j$ are measured uncertainties $h_{i}$ and $h_j$ respectively. 
\\\\
The general cases considered are:
\\ \textit{Case 1:} When $T_{S_i}$ is leaked and $S_i$ needs to be kept secret.
\\ \textit{Case 2:} When $T_{S_i}$ is leaked and $S_i$ and/or $S_j$ needs to be kept secret.
\\\\
The admissible rate region for each case is defined as follows:
\\ \textit{Definition 1a:} ($R_i$, $R_{ki}$, $h_{i}$) is admissible for case 1 if there exists a code ($F_{E_{i}}$, $F_{D}$) such that \eqref{ms_cond1} - \eqref{ms_cond7} hold for any $\epsilon \rightarrow 0$ and sufficiently large $K$.
\\ \textit{Definition 1b:} ($R_i$, $R_{ki}$, $R_j$, $R_{kj}$, $h_{j}$) is admissible for case 2 if there exists a code ($F_{E_{i}}$, $F_{D}$) such that \eqref{ms_cond1} - \eqref{ms_cond3} and \eqref{ms_cond8} hold for any $\epsilon \rightarrow 0$ and sufficiently large $K$.
\\ \textit{Definition 2:} The admissible rate regions are defined as:

\begin{eqnarray}
\mathcal{R}(h_{i}) = \{(R_i, R_{ki}):			\nonumber
\\(R_i, R_{ki}, h_{i} ) \text{ is admissible for case 1} \}
\end{eqnarray}

\begin{eqnarray}
\mathcal{R}(h_{i}, h_{j}) = \{(R_i, R_{ki}, R_j, R_{kj}):			\nonumber
\\(R_i, R_{ki}, R_j, R_{kj}, h_{j} ) \text{ is admissible for case 2} \}
\end{eqnarray}

The theorems developed for these regions follow:

\textit{Theorem 6:} For $0 \le h_{i} \le I(S_i;S_n|S_n^c)$,
\begin{eqnarray}
&& \mathcal{R}_1(h_{i}) = \{(R_i, R_{ki}): 		\nonumber
\\ && R_i \geq H(S_i), 				\nonumber
\\ && R_{k_i} \geq I({\bf S}_t|{\bf S}_t^c) \}			
\label{theorem6}
\end{eqnarray}

\textit{Theorem 7:} For $0 \le h_{j} \le H(S_i, S_j)$,
\begin{eqnarray}
&& \mathcal{R}_2(h_{i}, h_{j}) = \{(R_i, R_{ki}, R_j, R_{kj}): 		\nonumber
\\ && R_i \geq H(S_i, S_j), R_j \geq H(S_i, S_j), 							\nonumber
\\ && R_{ki} \geq I(S_i;S_j) \text{ and } R_{kj} \geq I(S_i;S_j) \}			
\label{theorem7}
\end{eqnarray}

The proofs for these theorems follow. The source information components are first identified. Assume the private portions of source $i$ and $j$ are given by $W_i$ and $W_j$ respectively.  

\begin{proof}[Theorem 6 proof]
Here, $R_i \geq H(S_i)$, $R_{ki} \geq I({\bf S}_t|{\bf S}_t^c)$. For the case where $h_i > I({\bf S}_t|{\bf S}_t^c)$, the definitions for $W_{CS_i}$, $W_i$ and $W_{ki}$ follow:

\begin{eqnarray}
W_{CS_i} = 2^{K I({\bf S}_t|{\bf S}_t^c)}
\label{theorem6_proof_41}
\end{eqnarray}

\begin{eqnarray}
W_i = (W_{Pi}, W_{kCi})
\label{theorem6_proof_43}
\end{eqnarray}

\begin{eqnarray}
W_{ki} = W_{Ci}
\label{theorem6_proof_44}
\end{eqnarray}

The keys and uncertainties are calculated as follows:

\begin{eqnarray}
\frac{1}{K} \log M_{i} & = & \frac{1}{K} (\log W_{S_i} + \log W_{CS_i}) \nonumber
\\ & \le & H(S_i|{\bf S}_{\backslash {S_i}}) + \frac{1}{K} W_{CS_i} + \epsilon_0 	\nonumber
\\ & = & H(S_i|{\bf S}_{\backslash {S_i}}) + I({\bf S}_t|{\bf S}_t^c)  + \epsilon_0 \nonumber 
\\ & = & \frac{1}{K} H(S_i) + \epsilon_0			\nonumber
\\ & \le & R_i + \epsilon_0
\label{theorem6_proof_45.0}
\end{eqnarray}

\begin{eqnarray}
&& \frac{1}{K} \log M_{ki} 		\nonumber
\\ & = & \frac{1}{K} \log W_{CS_i}	\nonumber
\\ & = & I({\bf S}_t|{\bf S}_t^c) 
\\ & \le & R_{ki} + \epsilon_0
\label{theorem6_proof_45.1}
\end{eqnarray}

\begin{eqnarray}
&& \frac{1}{K} H(S_i|W_{Pi}, W_{Ci}) 		\nonumber
\\ & \geq & \frac{1}{K} H(S_i) - \epsilon_0^{'} 		\nonumber
\\ & = & h_i - \epsilon_0^{'}		
\label{theorem6_proof_47}
\end{eqnarray}

From \eqref{theorem6_proof_45.0} - \eqref{theorem6_proof_47}, ($R_i$, $R_{ki}$, $h_i$) is admissible for $h_i > I({\bf S}_t|{\bf S}_t^c))$. We now consider the case where $h_i \le I({\bf S}_t|{\bf S}_t^c))$ and define $W_{CS_i}$, $W_i$ and $W_{ki}$ as follows:

\begin{eqnarray}
W_{CS_i} = 2^{K I({\bf S}_t|{\bf S}_t^c)}
\label{theorem6_proof_49}
\end{eqnarray}

\begin{eqnarray}
W_{i} = (W_{Pi}, W_{kCi})
\label{theorem6_proof_53}
\end{eqnarray}

\begin{eqnarray}
W_{ki} = W_{Ci}
\label{theorem6_proof_54}
\end{eqnarray}

The keys and uncertainties are calculated as follows:
\begin{eqnarray}
&& \frac{1}{K} \log M_{ki} 		\nonumber
\\ & = & \frac{1}{K} \log W_{CS_i}	\nonumber
\\ & = & I({\bf S}_t|{\bf S}_t^c)) 
\\ & \le & R_{ki} + \epsilon_0
\label{theorem6_proof_55}
\end{eqnarray}

\begin{eqnarray}
&& \frac{1}{K} H(S_i|W_{Pi}, W_{Ci}) 		\nonumber
\\ & \geq & \frac{1}{K} H(S_i|W_{Ci}) + I({\bf S}_t|{\bf S}_t^c) - \epsilon_0^{'} 		\nonumber
\\ & = & H(S_i) - \epsilon_0^{'} \nonumber
\\ & = & h_i - \epsilon_0^{'}
\label{theorem6_proof_57}
\end{eqnarray}

From \eqref{theorem6_proof_55} - \eqref{theorem6_proof_57} it is seen that ($R_i$, $R_{ki}$, $h_i$) is admissible for $h_i \le I({\bf S}_t|{\bf S}_t^c))$.
\end{proof}

Theorem 7 is proven in a similar manner. 

\begin{proof}[Theorem 7 proof]
Here, $R_i \geq H(S_i, S_j)$, $R_j \geq H(S_i, S_j)$, $R_{ki} \geq I(S_i;S_j)$ and $R_{kj} \geq I(S_i;S_j)$. For the case where $h_j \le H(S_i;S_j)$, the definitions for $W_{CS_i}$, $M_{Cj}$ $W_i$, $W_{ki}$, $W_j$ and $W_{kj}$ follow:

\begin{eqnarray}
W_{CS_i} = 2^{K I({\bf S}_t|{\bf S}_t^c)}
\label{theorem7_proof_40.9}
\end{eqnarray}

\begin{eqnarray}
M_{Cj} = 2^{K I({\bf S}_t|{\bf S}_t^c)}
\label{theorem7_proof_40.10}
\end{eqnarray}

\begin{eqnarray}
W_i = (W_{Pi}, W_{kCi})
\label{theorem7_proof_41}
\end{eqnarray}

\begin{eqnarray}
W_{ki} = W_{Ci}
\label{theorem7_proof_42}
\end{eqnarray}

\begin{eqnarray}
W_j = (W_{Pj}, W_{kCj})
\label{theorem7_proof_43}
\end{eqnarray}

\begin{eqnarray}
W_{kj} = W_{Cj}
\label{theorem7_proof_44}
\end{eqnarray}

The keys and uncertainties are calculated as follows:

\begin{eqnarray}
\frac{1}{K} \log M_{i} & = & \frac{1}{K} (\log W_{S_i} + \log W_{CS_i}) \nonumber
\\ & \le & \frac{1}{K} H(S_i|{\bf S}_{\backslash S_i}) + \frac{1}{K} W_{CS_i} + \epsilon_0 	\nonumber
\\ & = & \frac{1}{K} H(S_i|{\bf S}_{\backslash S_i}) + I({\bf S}_t|{\bf S}_t^c)  + \epsilon_0 	
\\ & = & \frac{1}{K} H(S_i) + \epsilon_0			\nonumber
\\ & \le & R_i + \epsilon_0
\label{theorem7_proof_45.0}
\end{eqnarray}

\begin{eqnarray}
\frac{1}{K} \log M_{j} & = & \frac{1}{K} (\log M_{Pj} + \log M_{Cj}) \nonumber
\\ & \le & \frac{1}{K} H(S_j|{\bf S}_{\backslash S_j}) + \frac{1}{K} M_{Cj} + \epsilon_0 	\nonumber
\\ & = & \frac{1}{K} H(S_j|{\bf S}_{\backslash S_j}) + I(S_j;{\bf S}_t|{\bf S}_t^c)  + \epsilon_0 	
\\ & = & \frac{1}{K} H(S_j) + \epsilon_0			\nonumber
\\ & \le & R_j + \epsilon_0
\label{theorem7_proof_45.1}
\end{eqnarray}

\begin{eqnarray}
&& \frac{1}{K} \log M_{ki} 		\nonumber
\\ & = & \frac{1}{K} \log W_{CS_i}	\nonumber
\\ & = & I({\bf S}_t|{\bf S}_t^c)
\\ & \le & R_{ki} + \epsilon_0
\label{theorem7_proof_46}
\end{eqnarray}

\begin{eqnarray}
&& \frac{1}{K} \log M_{kj} 		\nonumber
\\ & = & \frac{1}{K} \log M_{Cj}	\nonumber
\\ & = & I(S_i; S_j) + \epsilon_0
\\ & \le & R_{ki} + \epsilon_0
\label{theorem7_proof_46.1}
\end{eqnarray}

\begin{eqnarray}
&& \frac{1}{K} H(S_j|W_{Pi}, W_{Ci}) 		\nonumber
\\ & \geq & H(S_j) - H(S_i) - \epsilon_0^{'} 		\nonumber
\\ & = & H(S_i, S_j) - H(S_i) - \epsilon_0^{'} \nonumber
\\ & \geq & h_j - H(S_i)
\\ & = & h_j - h_i - \epsilon_0^{'}		
\label{theorem7_proof_47}
\end{eqnarray}

From \eqref{theorem7_proof_45.0} - \eqref{theorem7_proof_47}, ($R_i$, $R_{ki}$, $R_j$, $R_{kj}$, $h_j$) is admissible for $h_j \le H(S_i, S_j)$. We now consider the case where $h_j > H(S_i,S_j)$, and define $W_{CS_i}$, $W_i$ and $W_{ki}$ as follows:

\begin{eqnarray}
W_{CS_i} = 2^{K I({\bf S}_t|{\bf S}_t^c)}
\label{theorem7_proof_40.91}
\end{eqnarray}

\begin{eqnarray}
M_{Cj} = 2^{K I({\bf S}_t|{\bf S}_t^c)}
\label{theorem7_proof_40.11}
\end{eqnarray}

\begin{eqnarray}
W_i = (W_{Pi}, W_{kCi})
\label{theorem7_proof_48}
\end{eqnarray}

\begin{eqnarray}
W_{ki} = W_{Ci}
\label{theorem7_proof_49}
\end{eqnarray}

\begin{eqnarray}
W_j = (W_{Pj}, W_{kCj})
\label{theorem7_proof_50}
\end{eqnarray}

\begin{eqnarray}
W_{kj} = W_{Cj}
\label{theorem7_proof_51}
\end{eqnarray}

The keys and uncertainties are calculated as follows:

\begin{eqnarray}
&& \frac{1}{K} \log M_{ki} 		\nonumber
\\ & = & \frac{1}{K} \log W_{CS_i}	\nonumber
\\ & = & I({\bf S}_t|{\bf S}_t^c)
\\ & \le & R_{ki} + \epsilon_0
\label{theorem7_proof_52}
\end{eqnarray}

\begin{eqnarray}
&& \frac{1}{K} \log M_{kj} 		\nonumber
\\ & \le & I(S_i; S_j) + \epsilon_0
\\ & \le & R_{ki} + \epsilon_0
\label{theorem7_proof_53}
\end{eqnarray}

\begin{eqnarray}
&& \frac{1}{K} H(S_j|W_{Pi}, W_{Ci}) 		\nonumber
\\ & \le & H(S_j) - H(S_i) + \epsilon_0^{'} 		\nonumber
\\ & = & H(S_i, S_j) - H(S_i) + \epsilon_0^{'} \nonumber
\\ & \le & h_j - H(S_i)
\\ & = & h_j - h_i - \epsilon_0^{'}		
\label{theorem7_proof_54}
\end{eqnarray}

From \eqref{theorem7_proof_52} - \eqref{theorem6_proof_54} it is seen that ($R_i$, $R_{ki}$, $R_j$, $R_{kj}$, $h_j$) is admissible for $h_j \le H(S_i,S_j)$.
\end{proof}

These theorems demonstrate the necessary rates required for perfect secrecy. The goal of the Shannon's cipher aspect was to reduce the key lengths. The masking method explained in this section is able to use common information as keys and therefore minimise the key rates for the general cases. 

The information leakage described in the Slepian-Wolf aspect indicates the common information that should be given added protection to ensure that even less information will be leaked. 
The new extended model presented here also incorporates a multiple correlated sources approach using Shannon's cipher system, which is more practical than looking at two sources.

\section{Comparison to other models}
The two correlated sources model across a channel with an eavesdropper is a more generalised approach of Yamamoto's~\cite{shannon1_yamamoto} model. If we were to combine the links into one link, we would have the same situation as per Yamamoto's ~\cite{shannon1_yamamoto}. From Section VI it is evident that the model can be implemented for multiple sources with Sahnnon's cipher system. Due to the unique scenario incorporating multiple sources and multiple links, the new model is more secure as private information and common information from other link/s are required for decoding.

Further, information at the sources may be more secure in the new model because if one source is compromised then only one source's information is known. In Yamamoto's ~\cite{shannon1_yamamoto} method both source's information is contained at one station and when that source is compromised then information about both sources are known. The information transmitted along the channels (i.e. the syndromes) do not have a fixed length as per Yamamoto's ~\cite{shannon1_yamamoto} method. Here, the syndrome length may vary depending on the encoding procedure and nature of Slepian-Wolf codes, which is another feature of this generalised model. 

The generalised model also has the advantage that varying amounts of the common information $V_{CX}$ and $V_{CY}$ (in the case of two sources) may be transmitted depending on the security of the transmission link and/or sources. For example, for two correlated sources, if $Y$'s channel is not secure we can specify that more of the common information is transmitted from $X$. In this way we're able to make better use of the transmission link's security. For Yamamoto's ~\cite{shannon1_yamamoto} method the common information was transmitted as one portion, $V_{C}$.

In this model, information from more than one link is required in order to determine the information for one source. This gives rise to added security as even if one link is wiretapped it is not possible to determine the contents of a particular source. This is attributed to the fact that this model has separate common information portions, which is different to Yamamoto's model. 

Another major feature is that private information can be hidden using common information. Here, common information produced by a source may be used to mask its private codeword thus saving on key length. The key allocation is specified by general rules presented in Section VI. The multiple correlated sources model presents a combination masking scheme where more than one common information is used to protect a private information, which is a practical approach. This is an added feature developed in order to protect the system.  This approach has not been considered in the other models mentioned in this section. 
 
The work by Yang \textit{et al.}~\cite{feedback_yang} uses the concept of side information to assist the decoder in determining the transmitted message. The side information could be considered to be a source and is related to this work when the side information is considered as correlated information. Similar work with side information that incorporates wiretappers, by Villard and Piantanida \cite{pablo_secure_multiterminal} and Villard \textit{et al.} \cite{pablo_secure_transmission_receivers} may be generalised in the sense that side information can be considered to be a source, however this new model is distinguishable as syndromes, which are independent of one another are transmitted across an error free channel in the new model. Further, to the author's knowledge Shannon's cipher system has not been incorporated into these models by Villard and Piantanida \cite{pablo_secure_multiterminal} and Villard \textit{et al.} \cite{pablo_secure_transmission_receivers}.

\section{Future work}
This work has room for expansion and future work. It would be interesting to consider the case where the channel capacity has certain constraints (according to the assumptions in Section VI the channel capacity is enough at all times). In the new model, the channels are either protected by keys or not however this is limited and a real case scenario where there are varying security levels for the channels is an avenue for future work. Another aspect for expansion is to investigate the allocation of common information as keys to minimize additional keys with links having varying security levels and limited capacity.

\section{Conclusion}
The information leakage for two correlated sources across a channel with an eavesdropper was initially considered. Knowing which components contribute most to information leakage aids in keeping the system more secure, as these terms can be made more secure. The information leakage for the two correlated source model was quantified and proven. Shannon's cipher system was also incorporated for this model and channel and key rates to achieve perfect secrecy have been provided. The two correlated sources model has been extended for the network scenario where we consider multiple sources transmitting information across multiple links. The information leakage for this extended model is detailed. The channel and key rates are also considered for the multiple correlated source model when Shannon's chipher system is implemented. A masking method is further presented to minimize key lengths and a combination masking method is presented to address its practical shortcoming.

\bibliographystyle{IEEEtran}	
\bibliography{bib}

\end{document}